\documentclass[12pt]{article}
\usepackage{amssymb}
\usepackage{amsmath}
\usepackage{amsfonts}
\usepackage{graphicx}
\usepackage{natbib}
\usepackage{setspace}
\usepackage{lscape}
\usepackage{geometry}
\usepackage{rotating}
\usepackage{caption}
\usepackage{version}
\usepackage{titletoc}
\usepackage{afterpage}
\usepackage{hyperref}
\usepackage{placeins}
\usepackage{comment}
\usepackage{booktabs}
\usepackage{array}
\usepackage{subcaption}
\usepackage{threeparttable}

\newcolumntype{P}[1]{>{\raggedright\arraybackslash}p{#1}}
\providecommand{\U}[1]{\protect\rule{.1in}{.1in}}

\allowdisplaybreaks
\makeatletter
\g@addto@macro\normalsize{
\setlength\abovedisplayskip{7pt}
\setlength\belowdisplayskip{7pt}
\setlength\abovedisplayshortskip{7pt}
\setlength\belowdisplayshortskip{7pt}
}
\let\OLDthebibliography\thebibliography
\renewcommand\thebibliography[1]{
  \OLDthebibliography{#1}
  \setlength{\parskip}{0pt}
  \setlength{\itemsep}{0pt plus 0.3ex}
}
\makeatother
\hypersetup{
colorlinks = true,
urlcolor = blue, linkcolor = blue, 
citecolor = blue }
\begin{document}

\title{{\Large Structural Econometric Estimation of the Basic Reproduction Number for Covid-19 Across U.S. States and Selected Countries\thanks{We wish to thank Alessandro Rebucci, Alexander Chudik, and Ron Smith for
their valuable comments. We also gratefully acknowledge Alexander Chudik for helpful discussions regarding his codes.}}}
\author{ Ida Johnsson\thanks{Chewy, Inc., and University of Southern California.} \and M. Hashem Pesaran\thanks{University of Southern California, USA, and Trinity College, Cambridge, UK.}
\and Cynthia Fan Yang\thanks{Corresponding author. Department of Economics, Florida State University, 281
Bellamy Building, Tallahassee, FL 32306, USA. Email: \href{mailto:cynthia.yang@fsu.edu}
{cynthia.yang@fsu.edu}.} }
\maketitle

\begin{abstract}
This paper proposes a structural econometric approach to estimating the basic reproduction number ($\mathcal{R}_{0}$) of Covid-19. This approach identifies $\mathcal{R}_{0}$ in a panel regression model by filtering out the effects of mitigating factors on disease diffusion and is easy to implement. We apply the method to data from 48 contiguous U.S. states and a diverse set of countries. Our results reveal a notable concentration of $\mathcal{R}_{0}$ estimates with an average value of 4.5. Through a counterfactual analysis, we highlight a significant underestimation of the $\mathcal{R}_{0}$ when mitigating factors are not appropriately accounted for.

\vspace{0.5cm}


\end{abstract}

\pagenumbering{gobble}

\newpage\pagenumbering{arabic} \setcounter{page}{1}

\section{Introduction}

Since the Covid-19 outbreak, a rapidly growing body of literature has been devoted to estimating its reproduction numbers, which are standard epidemiological metrics used to quantify the rate at which the disease spreads at the initial and later stages of the epidemic. Obtaining accurate estimates of the reproduction numbers is critical in understanding the transmissibility of infectious agents and formulating public health responses. Different formal definitions of reproduction numbers have been proposed in the literature. One of the most widely adopted metrics is the basic reproduction number, denoted by $\mathcal{R}_{0}$, which is defined as "the average number of secondary cases produced by one infected individual during the infected individual's entire infectious period assuming a fully susceptible population" \citep{DelValle2013}. An essential assumption of this definition is that the population must be entirely susceptible without any immunity or interventions. In other words, the value of $\mathcal{R}_{0}$ represents the maximum rate at which an epidemic can spread \textit{in the absence of any mitigating actions}, whether voluntary or mandatory, such as improved personal hygiene, mask-wearing, social distancing, isolation, or vaccination. In practice, the day-to-day evolution of the epidemic is governed by public health directives and individual behavior, as well as the internal dynamics of the epidemic, which on its own eventually slows down its rate of spread as the number of susceptible individuals declines (due to immunity, death, or mitigation practices). A measure that quantifies the time-varying transmissibility during the life of the epidemic is the effective reproduction number, $\mathcal{R}_{e,t}$.\footnote{The effective reproduction number is the expected number of secondary cases produced by one infected individual in a population that includes both susceptible and non-susceptible individuals at time $t$ (assuming that the conditions remained the same after time $t$). The effective reproduction number differs from another time-varying metric, the case reproduction number \citep{Fraser2007}. The latter is defined as the average number of secondary cases that a primary case actually infects at time $t$, and thus its estimation can only be undertaken retrospectively.} The focus of the present study is on the structural estimation of $\mathcal{R}_{0}$ for Covid-19 across U.S. states and countries.

Obtaining a reliable estimate of $\mathcal{R}_{0}$, although seemingly straightforward, can be quite challenging in practice. This is largely due to the difficulty of accurately counting early infections during an epidemic, especially for newly emerging pathogens like SARS-CoV-2. Even if public health officials may strive to establish surveillance systems promptly, the data quality during the initial stages of the outbreak is inevitably poor. Additionally, many existing time-series estimation methods encounter difficulties in identifying an appropriate sample period to estimate $\mathcal{R}_{0}$. Unlike $\mathcal{R}_{e,t}$, which varies over time and is affected by mitigation policies, to estimate $\mathcal{R}_{0}$ it is important to focus on the initial stages of the epidemic when the infections are spreading exponentially, without impediments. If the selected sample is too long, it will likely include periods of reduced social interactions due to mandatory and/or voluntary behavioral changes, which slow down the spread of the infection and result in a downward bias in the $ \mathcal{R}_{0}$ estimates. On the other hand, if the sample selected at the initial stages of the epidemic is too short, the under-recording of infected cases can be particularly serious, resulting in an underestimation of the disease's transmission rate and again leading to the underestimation of $ \mathcal{R}_{0}$.

The availability of Covid-19 data from diverse populations and locations presents an unprecedented opportunity for researchers to compare the $ \mathcal{R}_{0}$ estimates using various models and estimation methods. Researchers around the world have reported a wide range of $\mathcal{R}_{0}$ values of Covid-19, as summarized in Table \ref{tab:lit_summary}. Some studies have highlighted the heterogeneity of $\mathcal{R}_{0}$ across different countries and regions \citep{Korolev2021, FernandezVillaverde2022}, while others find evidence of similarity \citep{Katul2020, Chudik2022}. Many of the existing estimates were obtained based on complex models that require large amounts of data and involve numerous assumptions made by the modeler. As it is rare to have high-quality data for all components of the model, researchers often have to calibrate multiple model parameters \citep{Delamater2019}. As our knowledge of the disease and the quality of data have improved over time, an increasing number of researchers have argued that earlier estimates of $\mathcal{R}_{0}$ were too conservative \citep{Katul2020, Ke2021}.

In the current paper, we estimate $\mathcal{R}_{0}$ of Covid-19 using a structural econometric approach, which allows us to identify $\mathcal{R}_{0} $ as the intercept in panel regressions of outcomes on a number of key factors across the contiguous states in the U.S. and a diverse range of countries. This structural method is grounded in the individual-based stochastic network model for epidemic diffusion recently developed by \cite {PY2022}, who established a moment condition for the evolution of the infections. \cite{Chudik2022} use the same moment condition but allow for time variations in the transmission rate, modeled as a function of various factors, including government responses (mandatory mitigation policies and economic support), precautionary behavioral changes such as voluntary social distancing, vaccinations, and virus mutations. These authors focus on nine European countries, which had similar starts at the outset of the epidemic in March 2020 but ended up with differing outcomes. They find robust statistically significant effects of reduced mobility, increased government economic support to comply with containment policies, vaccination, and virus mutations on the transmission rate of Covid-19.

We employ the structural approach proposed by \cite{Chudik2022}, CPR hereafter, and estimate $\mathcal{R}_{0}$ as the intercept in panel data regressions where suitably transformed case numbers are explained in terms of the mitigating factors identified in the CPR study. Such an estimate can be viewed as a counterfactual outcome that could have resulted if none of the public health policies had been enacted. This method has several advantages. First, by controlling for a number of mitigation measures, we are able to estimate $\mathcal{R}_{0}$ using relatively long samples, thus avoiding the difficulties of selecting an early sample. Second, the model easily accommodates the differences in the initial start dates of the Covid-19 outbreaks across countries (or regions) by using an unbalanced panel data model. Third, country-specific $\mathcal{R}_{0}$ can be estimated by least squares under very weak assumptions and only requires the regressors in the panel data model to be weakly exogenous. We also use lagged values of the regressors to avoid simultaneity bias, and we take account of error serial correlation when computing the standard errors to deal with the implications of using overlapping seven-day moving averages, a practice routinely employed in the literature to deal with uneven recording of infected cases over the different days of the week. Fourth, our estimation of $\mathcal{R}_{0}$ requires only Covid-19 case data and an assumed value for the recovery rate, $\gamma $. This approach serves as a useful complement to other methods that rely on mortality data. Instead of estimating the recovery rate, we set it \textit{a priori} using clinical evidence with a high degree of reliability. Fifth, we allow for the under-reporting of infected cases, which is known to vary across different phases of the epidemic.

We find remarkably similar estimates of $\mathcal{R}_{0}$ across 48 contiguous U.S. states, averaging at 4.7 with a range of 4.2 to 4.9 (4.5 to 5.3) under the assumption of a lower (higher) degree of under-reporting. In addition, for a selection of diverse nations, $\mathcal{R}_{0}$ remains within a narrow range of 3.4 to 5 across all scenarios considered, with an average value of 4.3. Overall, these estimates align well with the recent findings in the literature. Furthermore, our counterfactual analysis underscores a substantial underestimation of $\mathcal{R}_{0}$ that results from neglecting the mitigating factors governing the time profile of the Covid-19 transmission rate.

The rest of the paper is organized as follows. Section \ref{sec:lit_review} reviews the related literature. Section \ref{sec:methodology} outlines the empirical methodology. Section \ref{sec:results} presents the main findings, and Section \ref{sec:conclusion} concludes. Additional tables and figures are provided in an online supplement.

\section{Related Literature} \label{sec:lit_review}

The estimation of the reproduction number for infectious diseases, including Covid-19, has been a topic of extensive research in the literature. In the interest of space, we will highlight a selection of recent contributions that are closely related to our study.

There are many different approaches to the estimation of $\mathcal{R}_{0}$ that can be broadly categorized into mathematical and statistical approaches \citep{White2020}. The mathematical approaches rely on constructing a theoretical model, which makes explicit hypotheses about the biological mechanisms driving the transmission rate and its dynamics. Such hypotheses range from simple representations of the time needed to complete some part of the disease process (such as the incubation period) to complex agent-based models with explicit modeling of social interactions \citep{Lessler2016}. In contrast, statistical methods derive estimators of $ \mathcal{R}_{0}$ directly from data using probabilistic models. They do not require specifying a disease-approximating mathematical model and can be implemented using standard statistical software. 

The most widely used mathematical models for disease transmission are the compartmental models, most notably the classical SIR model that divides a population into three compartments (susceptible, infected, and recovered) and its variants, such as the SEIR and SIRD models (with an additional exposed or deceased compartment, respectively). \cite{Tang2020a} provide a recent review of the compartmental infectious disease models. Many researchers have made efforts to infer $\mathcal{R}_{0}$ of Covid-19 from different mathematical models and adopted various frequentist methods to estimate or calibrate the model parameters \citep[e.g.,][]{Buckman2020, Katul2020, Ke2021, Korolev2021, FernandezVillaverde2022}. Another strand of literature estimates similar compartmental models using Bayesian techniques \citep[e.g.,][]{Wu2020, Arias2023, Atkeson2023}.\footnote{Some authors categorize this strand of methods separately as stochastic approaches or stochastic Markov Chain Monte Carlo (MCMC) methods.}

The statistical methods typically require an estimate of the generation time, which is the time lag between infections in primary and secondary cases \citep{Obadia2012}. A frequently used estimator of $\mathcal{R}_{0}$ was proposed by \cite{Wallinga2007}, who show how to estimate $\mathcal{R} _{0}$ from the estimated exponential growth rate using data on infected cases during the \textit{early phase} of an outbreak and the moment generating function of the generation time distribution. Since the time lag between all infectee/infector pairs is not directly observable, the generation time distribution in practice is often substituted with the serial interval distribution that measures the time between symptoms onset. Besides the method of statistical exponential growth, other statistical approaches to estimating $\mathcal{R}_{0}$ include sequential Bayesian \citep{Bettencourt2008} and maximum likelihood estimations \citep{White2008}. See \cite{White2020} for a recent review of the statistical estimation of the reproduction number. Many researchers have adopted the statistical exponential growth model to estimate the $\mathcal{R}_{0}$ of Covid-19 using early case data for China \citep[e.g.,][]{Li2020, Liu2020a, Sanche2020, Zhao2020}.

\begin{table}[thb!]
    \centering
    \caption{Reported $\mathcal{R}_0$ estimates for Covid-19 from selected studies}
    \renewcommand{\arraystretch}{1.2}
    {\footnotesize
    \begin{tabular}{p{0.2\textwidth}P{0.15\textwidth}P{0.21\textwidth}P{0.39\textwidth}}
    \toprule
        Study & Locations & Methods & $\mathcal{R}_0$ estimates \\
        \midrule
        \cite{Arias2023} & Belgium & SIRD model, MCMC & Slightly above 4 \\
        \cite{Buckman2020} & Brazil, China, Italy, US & SEIR model & Range from 4.8 to 11.4 (US=9.7) \\
        \cite{Chudik2022} & 9 Euro countries & Structural econometric approach & Range from 5.01 to 5.51 \\
        \cite{FernandezVillaverde2022} & Various cities, states, countries & SIRD model & Range from 1.5 to 2.4 (US=2.02) \\
        \cite{Katul2020} & 57 countries & SIR model & 4.5 \\
        \cite{Ke2021} & US and 8 Euro countries & SEIR model & Range from 3.6 and 6.4 (US=5.9 [4.7--7.5]) \\
        \cite{Korolev2021} & California, Japan, US & SEIRD model & Range from 1.69 to 19.55 \\
        \cite{Li2020} & China & Statistical exponential growth model & 2.2 [1.4--3.9] \\
        \cite{Liu2020a} & China & Statistical exponential growth model & 2.90 [2.32--3.63] \\
        & & Statistical maximum likelihood estimation & 2.92 [2.28--3.67] \\
        \cite{Locatelli2021} & 15 Euro countries & Statistical exponential growth model & 2.2 [1.9--2.6] \\
        \cite{Sanche2020} & China & Statistical exponential growth model & 5.7 [3.8--8.9] \\
        \cite{Wu2020} & China & SEIR model, MCMC & 2.68 [2.47--2.86] \\
        \cite{Zhao2020} & China & Statistical exponential growth model & Range from 2.24 [1.96--2.55] to 3.58 [2.89--4.39] \\
     \bottomrule
    \end{tabular}
    }
    \label{tab:lit_summary}  
    \vspace{-0.2cm}\flushleft\footnotesize{Note: The 95\% confidence intervals are in brackets.} 
\end{table}

To facilitate comparisons, Table \ref{tab:lit_summary} provides a summary of reported $\mathcal{R}_{0}$ estimates and respective methods from selected studies. This table complements earlier summaries that focus on the $ \mathcal{R}_{0}$ estimates for China \citep{Alimohamadi2020, Liu2020} by showcasing the more recent estimates for various locations, with special emphasis on the U.S. estimates.\footnote{\cite{Alimohamadi2020} found that the estimates of $\mathcal{R}_{0}$ based on 23 studies using Chinese data ranged from 1.90 to 6.49, with a mean value of 3.38. \cite{Liu2020} surveyed 12 studies reporting a mean $\mathcal{R}_{0}$ estimate for China of 3.28, with a range of 1.4 to 6.49.}

\section{Empirical Methodology} \label{sec:methodology}

Our econometric approach follows from a simplified moment condition of an individual-based stochastic network SIR model developed by \cite{PY2022}. Assuming a homogeneous population,\footnote{\cite{PY2022} also considered a heterogeneous population segmented into multiple groups that can be based on demographic and socioeconomic factors as well as contact locations/schedules. Using simulations, they find that when $n$ is sufficiently large and the rate of infection is reasonably rapid, there are little differences in the time profiles of infected cases at the aggregate level whether one uses a single-group or multi-group network model, which partly justify our use of single-group moment conditions for estimation of $\mathcal{R}_{0}$.} they derive the following moment condition in $c_{t}$ (per capita number of cumulative cases on day $t$) and $i_{t}$ (per capita number of active infected cases on day $t$): 
\begin{equation}
E\left( \frac{1-c_{t+1}}{1-c_{t}}|i_{t}\right) =e^{-\beta _{t}i_{t}}+O\left(
n^{-1}\right) ,  \label{eq:moment}
\end{equation}
where $\gamma $ is the recovery rate (per day) and is assumed to be time-invariant; $\beta _{t}$ denotes the transmission rate, which can vary over time; and $n$ stands for the population size. Under this setup, $i_{t}$ can be computed from current and past values of $c_{t}$ for a given choice of $\gamma$:
\begin{equation*}
i_{t}=c_{t}-\gamma \sum\limits_{\ell =0}^{\infty }(1-\gamma )^{\ell}c_{t-\ell -1}, 
\end{equation*} 
or recursively, using $i_{t}=(1-\gamma )i_{t-1}+\Delta c_{t},$ where $\Delta c_{t}=c_{t}-c_{t-1}$ denotes per-capita daily new cases, with $c_{t}=i_{t}=0$ , for $t\leq 0$  (before the start of the epidemic).

Since $n$ is large and $\beta _{t}i_{t}$ is small, Eq. (\ref{eq:moment}) implies that $\beta _{t}$ can be approximated by $-i_{t}^{-1}\ln \left( \frac{1-c_{j,t+1}}{1-c_{jt}} \right)$. To account for additional factors that may influence disease transmission, CPR model the time variations of $\beta _{t}$ in terms of a set of variables measuring mandatory or voluntary social distancing, economic support, vaccine uptake, and virus mutations. We adopt CPR's strategy and estimate the following threshold panel data model: 
\begin{equation}
\beta _{jt}/\gamma =\frac{-\ln \left( \frac{1-c_{j,t+1}}{1-c_{jt}}\right) }{
\gamma i_{jt}}=\alpha _{j}+\boldsymbol{\psi }^{\prime }\boldsymbol{x}
_{j,t-p}+\kappa I(\Delta c_{j,t-p}>\tau )+u_{j,t+1},  \label{eq:estimating}
\end{equation} 
in $n$ states or countries indexed by $j=1,2,\ldots ,n,$ over the days $ t=1,2,\ldots ,T$. The dependent variable, $\beta _{jt}/\gamma $, measures the transmission rate in country $j$ scaled by the recovery rate, $\gamma $. One can also interpret $1/{\gamma }$ as the mean infectious period. Based on the clinical evidence for Covid-19 and in line with quarantine policies, we set $\gamma =1/14$, which is also a commonly adopted value in previous studies.\footnote{If data on the recovered cases are available, one can also estimate the recovery rate using, for example, the recovery moment condition outlined in \cite{PY2022}. However, we have opted to calibrate the recovery rate value based on clinical information due to its greater reliability, and this is the only parameter we need to calibrate.}

Turning to the right-hand side variables of Eq. (\ref{eq:estimating}), $ \boldsymbol{x}_{j,t-p}$ is a vector of explanatory variables lagged $p$ days, and $u_{j,t+1}$ are the idiosyncratic errors. The precautionary behavior is captured by the term $I(\Delta c_{j,t-p}>\tau )$, where $I(\cdot )$ is the indicator function equal to one if $\Delta c_{j,t-p}>\tau $ and zero otherwise. $\Delta c_{j,t-p}$ is the (seven-day moving average of) reported daily new cases per 100,000 people also lagged $p$ periods. We chose this as the threshold variable since it has been the most closely monitored metric on the spread of Covid-19 in media reports worldwide. $\tau  $ is a threshold parameter and, for simplicity, is assumed to be the same across countries. Intuitively, higher values of $\Delta c_{j,t-p}$ imply greater perceived risk; individuals are likely to practice voluntary social distancing more vigorously if $\Delta c_{j,t-p}$ exceeds the threshold value  $\tau $.

The parameters to be estimated include $\alpha _{j}$, $\boldsymbol{\psi }$, $ \kappa $, and $\tau $. In the present study, the key structural parameter of interest is the state- or country-specific intercept, $\alpha _{j}$, which we use as the estimator of the mean basic production number, $\mathcal{R}_{0} $, for state (or country) $j$.\footnote{In contrast to CPR who also considered a common intercept term for the nine European countries in their study, we focus on the fixed-effects (FE) estimation because our study covers a larger set of heterogeneous countries and states in the U.S.} This is valid because in the absence of any mitigating interventions, whether mandatory and/or voluntary, both $\mathbf{x }_{j,t-p}$ and $I(\Delta c_{j,t-p}>\tau )$ would equal to zero, and then $ \beta _{jt}/\gamma =\beta _{j0}/\gamma $ for all $t$. Hence, the intercept serves as an estimate of $\beta _{j0}/\gamma $, which is indeed the same as $ \mathcal{R}_{0}$ in the standard SIR models. To identify $\alpha _{j}$, however, it is important that we take into account the mitigating effects of behavioral changes in response to the spread of the epidemic, uptake of vaccination, and possible mutations of the virus. It is also worth noting that $c_{jt}$ is very small at the onset of the outbreak, and hence $-\ln \left( \frac{1-c_{j,t+1}}{1-c_{jt}}\right) \approx \Delta c_{j,t+1}$. This approximation, coupled with the fact that $\mathbf{x}_{j,t-p}$ and $I(\Delta c_{j,t-p}>\tau )$ are both equal to zero at the initial stages of the epidemic, makes Eq. (\ref{eq:estimating}) consistent with the basic idea of estimating $\mathcal{R}_{0}$ by the average growth of infections per active cases over its infectious period.

We estimate model (\ref{eq:estimating}) over two sample periods. The first sample covers the period from March 6, 2020, to January 31, 2021, prior to the roll-out of Covid-19 vaccinations. The second sample extends the first one to November 30, 2021. We will refer to these two samples as the pre-vaccination sample and the full sample, respectively. Both samples are unbalanced since the dates of the outbreaks of Covid-19 and the available case data differ across U.S. states and countries.

Apart from the threshold indicator variable, we also include policy stringency and economic support indices as mitigating factors for both sample periods. For the full sample, we include two additional factors, namely the share of the population that is fully vaccinated and the share of the Delta variant of Covid-19. The rationale behind including the latter factor stems from its greater transmissibility as compared to the earlier variants.

For the U.S. state-level regressions, we also include the interaction between the economic support index and a dummy variable for a Republican governor. This allows us to account for the varied effects of economic policies on the spread of Covid-19 across states with different political leanings, since the economic support index does not include support to firms or take into account the total fiscal value of economic support provided \citep{Hale2020}.

We utilized data from multiple sources to conduct our analysis. The policy stringency and economic support indices were retrieved from the Oxford Covid-19 Government Response Tracker (OxCGRT) project.\footnote{Available at \url{https://github.com/OxCGRT/covid-policy-dataset}.} The data on Delta variants were obtained from CoVariants.org \citep{Hodcroft2021}. Our World in Data was the source of the data on Covid-19 vaccinations as well as the country-level Covid case numbers.\footnote{Available at \url{https://github.com/owid/covid-19-data/tree/master/public/data}.} The Centers for Disease Control and Prevention's Covid data tracker was used to obtain state-by-state Covid cases in the U.S.\footnote{Downloaded from \url{https://data.cdc.gov/api/views/9mfq-cb36/rows.csv?accessType=DOWNLOAD} (last accessed: May 2023).}

To reduce the impact of within-week seasonality in the reported daily cases due to delayed reporting or reduced testing on weekends, we follow the standard practice and take seven-day moving averages of the reported data before estimation. Another important data issue that needs to be addressed is the under-reporting of confirmed cases. This occurs mainly due to asymptomatic infections and lack of testing, especially during the early stages of an outbreak. In the epidemiological literature, the magnitude of under-reporting is often measured by the multiplication factor (MF), which is defined as the ratio of true to reported cases \citep[e.g.,][]{Gibbons2014}. For Covid-19, with increased testing we would expect MF to fall over time, and simulations carried out by \citep{PY2022} support this. To account for this declining under-reporting, we consider two different specifications of the MF values: one where the MF linearly declines from 5 to 2, and another where it declines from 8 to 2.5.

We compute three types of standard errors in all estimations. The first type is the ``usual" standard error assuming no cross-sectional or serial correlation in the errors. The second type, labeled as ``robust1" in the tables, is the Newey-West type heteroskedasticity and autocorrelation consistent standard error \citep{NeweyWest1987}. The third type, labeled ``robust2", is the \cite{DriscollKraay1998} standard error, which accounts for both cross-sectional and serial correlation. For the latter two robust standard errors, we choose the truncation lag as the integer part of $T^{1/3}$, where $T$ denotes the longest time span of the unbalanced panel.

\section{The Main Findings} \label{sec:results}

This section presents the estimates of $\mathcal{R}_{0}$ for the U.S. states and 19 countries. In assessing the estimates' statistical significance, we will focus on the most conservative standard errors that are robust to both error cross-sectional and serial correlation. Estimates for the mitigating covariates and their standard errors are provided in the online supplement.

\subsection{Estimates of $\mathcal{R}_{0}$ for U.S. states}

\begin{table}[!hp]
\caption{Estimates of $\mathcal{R}_0$ across 48 contiguous U.S. states}
\centering 
\vspace{-0.3cm}
\begin{footnotesize}
\begin{tabular}{lccccc}\toprule 
&\multicolumn{2}{c}{\bf Pre-vaccination sample}&&\multicolumn{2}{c}{\bf Full sample} \\
&\multicolumn{2}{c}{\bf Ending Jan 31, 2021}&&\multicolumn{2}{c}{\bf Ending Nov 30, 2021} \\
\cline{2-3} \cline{3-3} \cline{5-6} \cline{6-6} 
& MF & MF && MF& MF \\
State& 5 to 2& 8 to 2.5 && 5 to 2 & 8 to 2.5 \\ \midrule
Alabama&4.29 [0.92]&4.60 [0.86]&&4.45 [0.81]&4.68 [0.76]\\ 
Arizona&4.62 [0.98]&4.92 [0.90]&&4.70 [0.83]&4.94 [0.78]\\ 
Arkansas&4.35 [0.97]&4.66 [0.90]&&4.50 [0.83]&4.76 [0.79]\\ 
California&4.77 [0.97]&5.08 [0.88]&&4.94 [0.90]&5.11 [0.87]\\ 
Colorado&4.62 [0.96]&4.92 [0.87]&&4.67 [0.85]&4.79 [0.80]\\ 
Connecticut&4.82 [1.02]&5.13 [0.92]&&4.79 [0.87]&4.91 [0.82]\\ 
Delaware&4.72 [1.01]&5.04 [0.92]&&4.75 [0.87]&4.93 [0.81]\\
District of Columbia&4.74 [1.09]&5.04 [0.98]&&4.68 [0.89]&4.77 [0.84]\\ 
Florida&4.41 [0.97]&4.73 [0.89]&&4.53 [0.83]&4.74 [0.78]\\ 
Georgia&4.41 [0.96]&4.76 [0.88]&&4.48 [0.83]&4.70 [0.78]\\ 
Idaho&4.36 [1.03]&4.63 [0.95]&&4.54 [0.85]&4.77 [0.80]\\ 
Illinois&4.68 [0.98]&5.00 [0.89]&&4.72 [0.84]&4.90 [0.79]\\ 
Indiana&4.34 [0.94]&4.66 [0.87]&&4.53 [0.84]&4.76 [0.79]\\ 
Iowa&4.29 [0.84]&4.73 [0.79]&&4.53 [0.80]&4.78 [0.76]\\ 
Kansas&4.65 [1.00]&4.95 [0.92]&&4.69 [0.84]&4.88 [0.79]\\ 
Kentucky&4.72 [1.06]&5.01 [0.96]&&4.73 [0.87]&4.92 [0.82]\\ 
Louisiana&4.70 [1.03]&5.01 [0.94]&&4.73 [0.89]&4.97 [0.84]\\ 
Maine&4.67 [1.15]&4.91 [1.04]&&4.71 [0.91]&4.74 [0.86]\\ 
Maryland&4.68 [1.01]&4.98 [0.92]&&4.66 [0.86]&4.73 [0.81]\\ 
Massachusetts&4.60 [1.05]&4.85 [0.96]&&4.79 [0.90]&4.95 [0.84]\\ 
Michigan&4.56 [1.05]&4.82 [0.96]&&4.71 [0.88]&4.86 [0.83]\\
Minnesota&4.57 [1.03]&4.85 [0.94]&&4.77 [0.87]&4.94 [0.82]\\ 
Mississippi&4.34 [0.95]&4.67 [0.89]&&4.44 [0.82]&4.67 [0.77]\\
Missouri&4.26 [0.94]&4.53 [0.87]&&4.46 [0.82]&4.64 [0.77]\\ 
Montana&4.27 [0.99]&4.58 [0.92]&&4.55 [0.84]&4.77 [0.79]\\ 
Nebraska&4.35 [0.96]&4.71 [0.89]&&4.51 [0.82]&4.77 [0.77]\\
Nevada&4.39 [1.01]&4.69 [0.93]&&4.55 [0.86]&4.78 [0.81]\\ 
New Hampshire&4.23 [0.96]&4.47 [0.89]&&4.49 [0.85]&4.56 [0.80]\\ 
New Jersey&4.73 [0.98]&5.07 [0.89]&&4.75 [0.85]&4.91 [0.80]\\ 
New Mexico&4.79 [1.23]&5.07 [1.11]&&4.87 [0.94]&5.06 [0.88]\\ 
New York&4.85 [1.01]&5.18 [0.91]&&4.66 [0.87]&4.70 [0.82]\\ 
North Carolina&4.69 [1.04]&4.96 [0.95]&&4.66 [0.87]&4.81 [0.82]\\ 
North Dakota&4.24 [0.89]&4.68 [0.84]&&4.67 [0.80]&5.17 [0.76]\\ 
Ohio&4.41 [1.05]&4.66 [0.96]&&4.56 [0.87]&4.72 [0.81]\\ 
Oklahoma&4.21 [0.87]&4.50 [0.82]&&4.48 [0.82]&4.72 [0.77]\\ 
Oregon&4.55 [0.97]&4.82 [0.89]&&4.60 [0.86]&4.62 [0.81]\\ 
Pennsylvania&4.64 [0.98]&4.92 [0.90]&&4.65 [0.84]&4.75 [0.79]\\ 
Rhode Island&4.88 [1.01]&5.29 [0.92]&&4.95 [0.88]&5.26 [0.83]\\ 
South Carolina&4.32 [0.92]&4.64 [0.85]&&4.51 [0.82]&4.75 [0.77]\\ 
South Dakota&4.25 [0.88]&4.65 [0.83]&&4.64 [0.81]&5.05 [0.76]\\ 
Tennessee&4.43 [1.03]&4.73 [0.95]&&4.59 [0.85]&4.92 [0.80]\\ 
Texas&4.45 [1.01]&4.73 [0.93]&&4.54 [0.85]&4.73 [0.80]\\ 
Utah&4.33 [0.96]&4.65 [0.89]&&4.55 [0.82]&4.87 [0.77]\\ 
Vermont&4.23 [1.05]&4.59 [0.97]&&4.45 [0.88]&4.47 [0.83]\\ 
Virginia&4.36 [0.98]&4.64 [0.91]&&4.48 [0.85]&4.59 [0.80]\\ 
Washington&4.65 [0.99]&4.94 [0.91]&&4.64 [0.88]&4.68 [0.83]\\ 
Wisconsin&4.57 [0.96]&4.89 [0.88]&&4.76 [0.85]&4.99 [0.80]\\ 
Wyoming&4.55 [1.02]&4.77 [0.94]&&4.58 [0.84]&4.69 [0.79]\\ 
\bottomrule\end{tabular}
\end{footnotesize}
\begin{tablenotes}
\scriptsize
\item
Notes: The pre-vaccination sample includes 14,996 observations for $N=48$ states, covering a time span of $T_{\min}=163$ to $T_{\max}=331$ days. The full sample comprises 29,531 observations, spanning $T_{\min}=457$  to $T_{\max}=634$  days. Both estimation samples are unbalanced at the beginning. Numbers in brackets are standard errors robust to serial correlation and cross-sectional correlation (robust2). The multiplication factor (MF) declines linearly from the high value to the low value. The estimates on the mitigating factors are provided in Table \ref{tab:state_regressors} in the online supplement.
\end{tablenotes}
\label{tab:state_R0}
\end{table}

We use the estimates of $\alpha _{j}$ from the panel threshold regressions, (\ref{eq:estimating}), after filtering out the effects of the mitigating factors, to obtain an estimate of $\mathcal{R}_{0}$ for the $j^{th}$ state. These estimates are summarized in Table \ref{tab:state_R0} for all 48 contiguous states in the U.S. We report two sets of estimates for each of the two sample periods that end on January 31, 2021, and November 30, 2021, respectively. We scale up all reported cases by $MF_{t}$ that declines linearly from 5 to 2 in one scenario and from 8 to 2.5 in another scenario.\footnote{The parameter estimates for the mitigating factors are provided in Table \ref {tab:state_regressors} of the online supplement.}

Overall, the results exhibit remarkable similarities across all states and sample periods, with all estimates being statistically highly significant, even when the most robust standard errors are used. The results are also reasonably robust to the choice of MF and the sample period. When using lower MF values, the estimates of $\mathcal{R}_{0}$ vary between 4.21 (Oklahoma) to 4.88 (Rhode Island) in the pre-vaccination sample, and between 4.44 (Mississippi) and 4.95 (Rhode Island) in the full sample. With higher MF values, the estimates rise slightly and vary between 4.47 (New Hampshire) to 5.29 (Rhode Island) in the case of the pre-vaccination sample, and 4.47 (Vermont) to 5.26 (Rhode Island) when we use the full sample.

\begin{figure}[htb!]
\caption{Histograms of $\mathcal{R}_0$ estimates across 48 contiguous U.S.
states}
\label{fig:hist_states_wide}\centering
    \begin{subfigure}{0.4\textwidth}
        \includegraphics[width=\linewidth]{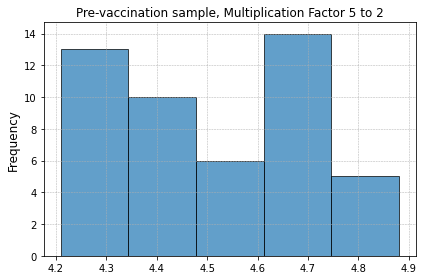}
    \end{subfigure} 
    \begin{subfigure}{0.4\textwidth}
        \includegraphics[width=\linewidth]{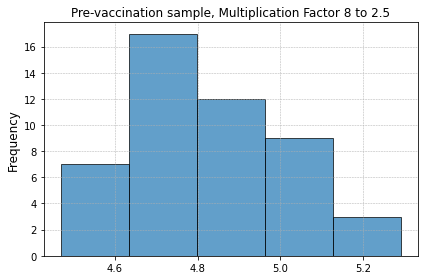}
    \end{subfigure} 
    \begin{subfigure}{0.4\textwidth}
        \includegraphics[width=\linewidth]{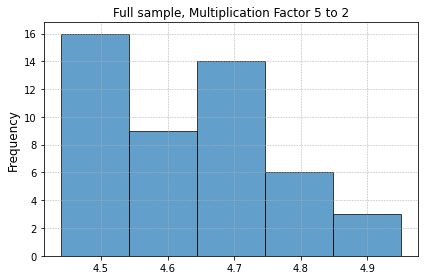}
    \end{subfigure}
    \begin{subfigure}{0.4\textwidth}
        \includegraphics[width=\linewidth]{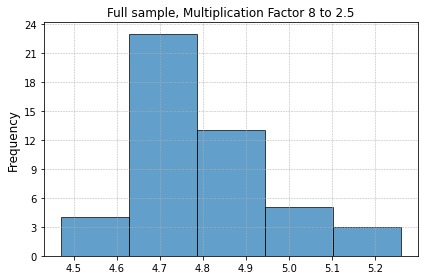}
    \end{subfigure} 
    \medskip 
    \flushleft{\footnotesize Notes: The pre-vaccination sample ends on January 31, 2021, and the full sample ends on November 30, 2021, for all  states. Both estimation samples are unbalanced at the beginning. }
\end{figure}

A closer inspection of results in Table \ref{tab:state_R0} reveals that the estimates of $\mathcal{R}_{0}$ tend to be slightly higher in states such as Rhode Island, California, Connecticut, and New Jersey, and lower in states such as Missouri, Oklahoma, and Alabama.\footnote{For a visual representation of the spatial distribution of the $\mathcal{R}_{0}$ estimates, refer to the hexagon maps presented in Figures \ref {fig:us_r0_subsample} and \ref{fig:us_r0_full_sample} of the online supplement.} Despite these differences, the estimates are very tightly clustered. This can be observed from the histograms in Figure \ref{fig:hist_states_wide}, which give the distribution of the estimates for the two sample periods and MF specifications. With lower MF values, the estimates are primarily clustered around 4.2 to 4.8. When the MF is higher, about 40 out of the 48 states have estimates concentrated between 4.6 and 5.1. To summarize, an average estimate (across states, periods, and MF values) of 4.7 seems to provide a reasonable summary number for the U.S.

 \subsection{Country-specific Estimates}

\begin{table}[!t]\caption{Estimates of $\mathcal{R}_0$ across 19 countries}
\centering 
\resizebox{12cm}{!}{
\begin{tabular}{lccccc}\toprule 
&\multicolumn{2}{c}{\bf Pre-vaccination sample}&&\multicolumn{2}{c}{\bf Full sample} \\
&\multicolumn{2}{c}{\bf Ending Jan 31, 2021}&&\multicolumn{2}{c}{\bf Ending Nov 30, 2021} \\
\cline{2-3} \cline{3-3} \cline{5-6} \cline{6-6} 
& MF & MF && MF& MF \\
Country& 5 to 2& 8 to 2.5 && 5 to 2 & 8 to 2.5 \\ \midrule
Argentina&5.01 [0.51]&5.04 [0.50]&&4.41 [0.61]&4.51 [0.59]\\ 
Australia&4.27 [0.49]&4.26 [0.48]&&3.92 [0.59]&3.92 [0.57]\\ 
Brazil&4.47 [0.45]&4.52 [0.44]&&4.08 [0.56]&4.17 [0.54]\\ 
Chile&4.85 [0.48]&4.90 [0.47]&&4.46 [0.61]&4.53 [0.59]\\ 
Colombia&4.83 [0.46]&4.86 [0.45]&&4.32 [0.57]&4.41 [0.55]\\ 
Ecuador&4.69 [0.48]&4.70 [0.48]&&4.22 [0.59]&4.24 [0.57]\\ 
Egypt&4.54 [0.41]&4.53 [0.41]&&3.95 [0.53]&3.97 [0.51]\\ 
France&4.58 [0.45]&4.61 [0.44]&&4.35 [0.57]&4.47 [0.55]\\ 
Germany&4.18 [0.44]&4.19 [0.43]&&4.08 [0.58]&4.11 [0.56]\\ 
Indonesia&4.08 [0.40]&4.08 [0.39]&&3.80 [0.54]&3.82 [0.52]\\ 
Japan&4.16 [0.36]&4.14 [0.36]&&3.97 [0.52]&3.97 [0.50]\\ 
Mexico&4.21 [0.40]&4.22 [0.40]&&3.83 [0.52]&3.85 [0.50]\\ 
Nigeria&3.88 [0.40]&3.88 [0.40]&&3.42 [0.49]&3.44 [0.48]\\ 
Peru&4.89 [0.45]&4.93 [0.45]&&4.31 [0.59]&4.38 [0.56]\\ 
Philippines&4.37 [0.43]&4.38 [0.42]&&3.88 [0.56]&3.90 [0.54]\\ 
South Korea&3.85 [0.37]&3.84 [0.37]&&3.78 [0.51]&3.77 [0.49]\\ 
Spain&4.83 [0.49]&4.87 [0.48]&&4.44 [0.61]&4.56 [0.58]\\ 
Thailand&4.64 [0.54]&4.63 [0.53]&&4.12 [0.57]&4.14 [0.55]\\ 
Turkey&4.69 [0.52]&4.70 [0.51]&&4.22 [0.60]&4.28 [0.58]\\ 
\bottomrule\end{tabular}}
\begin{tablenotes}
\footnotesize
\item
Notes: The pre-vaccination sample includes 5,928 observations for 19 countries, covering a time span of $T_{\min}=140$  to $T_{\max}=329$  days. The full sample comprises 11,661 observations for the same 19 countries, spanning $T_{\min}=443$  to $T_{\max}=632$  days. Both estimation samples are unbalanced at the beginning. Numbers in brackets are standard errors robust to serial correlation and cross-sectional correlation (robust2). The multiplication factor declines linearly from the high value to the low value. The estimates on the mitigating factors are provided in Table \ref{tab:country_regressors} in the online supplement.
\end{tablenotes}
\label{tab:country_R0}
\end{table}

\begin{figure}[htb!]
    \caption{Histograms of $\mathcal{R}_0$ estimates across 19 countries}
    \label{fig:hist_international_wide}\centering
    \begin{subfigure}{0.4\textwidth}
        \includegraphics[width=\linewidth]{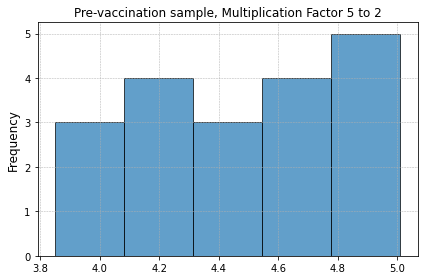}
    \end{subfigure} 
    \begin{subfigure}{0.4\textwidth}
        \includegraphics[width=\linewidth]{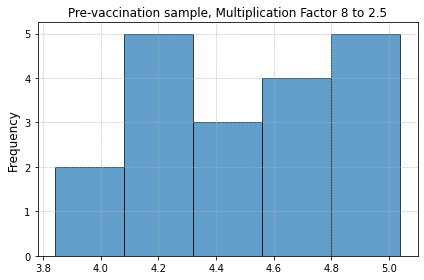}
    \end{subfigure} 
    \begin{subfigure}{0.4\textwidth}
        \includegraphics[width=\linewidth]{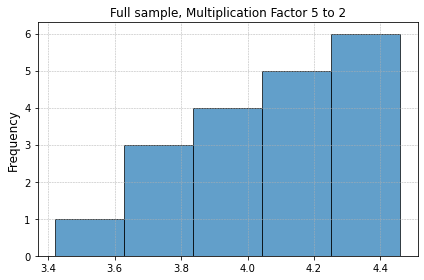}
    \end{subfigure}
    \begin{subfigure}{0.4\textwidth}
        \includegraphics[width=\linewidth]{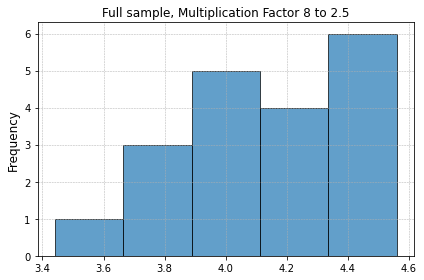}
    \end{subfigure} 
    \medskip 
    \flushleft{\footnotesize Notes: The pre-vaccination sample ends on January 31, 2021, and the full sample ends on November 30, 2021, for all states. Both estimation samples are unbalanced at the beginning. }
\end{figure}

Table \ref{tab:country_R0} reports the estimates of $\mathcal{R}_{0}$ for 19 countries.\footnote{The associated estimation results for the mitigating factors of the country regressions are provided in Table \ref{tab:country_regressors} in the online supplement, where we also present the $\mathcal{R}_{0}$ estimates in descending bar charts in Figures \ref{fig:int_r0_subsample} and \ref {fig:int_r0_full_sample}.} Similar to our findings for the U.S. states, the estimates fall within a narrow range. Moreover, the results are fairly robust to the choice of MF, with higher MF values only marginally increasing the $\mathcal{R}_{0}$ estimates. Argentina, Chile, Peru, and Spain are found to have the highest average $\mathcal{R}_{0}$ values, while Nigeria, South Korea, and Indonesia have the lowest average $\mathcal{R}_{0}$ values. Specifically, for the pre-vaccination period, the estimates range from 3.85 (South Korea) to 5.01 (Argentina) under the lower MF values, and 3.84 (South Korea) to 5.04 (Argentina) under the higher MF values. The full-sample estimates lie between 3.42 for Nigeria and 4.46 for Chile (3.44 for Nigeria and 4.56 for Spain) in the low (high) MF scenarios. The slight differences in estimated $\mathcal{R}_{0}$ values across countries might be associated with varying degrees of under-reporting. To further examine the distribution of the estimates, Figure \ref{fig:hist_international_wide} displays the histograms for each sample period and MF specification. We see that the estimates from the pre-vaccination sample are relatively evenly distributed within the range, whereas about 15 out of the 19 countries have full sample estimates falling in the range of 3.8 to 4.5. In sum, the average estimate of $ \mathcal{R}_{0}$ for 19 countries across both sample periods and MF values is around 4.3, which is quite close to the average U.S. estimate of 4.7. Overall, both our U.S. states and international estimates align with the recent literature and suggest that earlier studies have underestimated $ \mathcal{R}_{0}$, as documented in Table \ref{tab:lit_summary}.

\subsection{Estimates of $\mathcal{R}_{0}$ without Mitigating Factors} \label{subsec:est_without_mitigating}

To demonstrate the importance of filtering out the effects of the mitigating factors in the estimation of $\mathcal{R}_{0}$, we estimated Eq. (\ref {eq:estimating}) without any of the mitigating factors  (or by setting both $ \mathbf{x}_{j,t-p}$ and $I(\Delta c_{j,t-p}>\tau )$ to zero). The state- and country-specific estimates are presented in Tables \ref{tab:state_R0_no_mit} and \ref{tab:county_R0_no_mit} in the online supplement. Evidently, the estimated $\mathcal{R}_{0}$'s are significantly biased downward when the mitigating factors are not accounted for. The average estimate across sample periods and MF specifications is only 1.5 (1.3) for U.S. states (19 countries). These results underscore the problem of omitted variable bias resulting from neglecting mitigating factors.

\section{Conclusions} \label{sec:conclusion}

In this paper, we propose a novel approach to the estimation of the basic reproduction number, $\mathcal{R}_{0}$, of Covid-19, and provide estimates for U.S. states and a selected number of countries. Our approach falls under the category of counterfactual causal analysis, where the focus is to filter out the effects of mitigating factors on the diffusion of the virus, and thus identify $\mathcal{R}_{0}$ when the values of the mitigating factors are set to zero in the counterfactual exercise. Our estimates of $\mathcal{R} _{0}$ turn out to be centered around $4.5$, clustered closely across U.S. states and a selected number of countries. Not allowing for the mitigating factors results in estimates of $\mathcal{R}_{0}$ that suffer from substantial downward bias.

While our estimation approach is relatively simple to implement and yields satisfactory estimates, it is subject to an important limitation. In order to identify and provide accurate estimates of $\mathcal{R}_{0}$, it requires the availability of reliable data on mitigating factors, ideally taking into account all such factors in the analysis. In practice, this means that the method might not be applicable at the very early stages of an epidemic. Nevertheless, we believe our approach offers a useful alternative to the existing methods for $\mathcal{R}_{{0}}$ estimation, which also face the challenge of obtaining reliable early samples that are not subject to any mitigating interventions.

 \onehalfspacing {\footnotesize  \bibliographystyle{chicago} \bibliography{covid_us} }

 \subsection*{Statements and Declarations}

 {\footnotesize The authors declare that no funds, grants, or other support were received during the preparation of this manuscript. The authors have no relevant financial or non-financial interests to disclose.}

 \clearpage \appendix \renewcommand{\thetable}{S.\arabic{table}} \setcounter{table}{0}  \renewcommand{\thefigure}{S.\arabic{figure}} \setcounter{figure}{0}  
 \pagenumbering{arabic}
 \renewcommand*{\thepage}{S\arabic{page}}

 \onehalfspacing

\begin{center}
\large\textbf{Online Supplement to }

\large\textbf{``Structural Econometric Estimation of the Basic Reproduction Number for Covid-19 Across U.S. States and Selected Countries"}
\end{center}

 \begin{center} Ida Johnsson \\[0pt] Chewy, Inc., and University of Southern California \medskip

 M. Hashem Pesaran \\[0pt] University of Southern California, USA, and Trinity College, Cambridge, UK \medskip

 Cynthia Fan Yang \\[0pt] Florida State University \bigskip

 \today \end{center}

\bigskip This online supplement provides the coefficient estimates for the mitigating factors in the panel threshold regressions and their summary statistics for the U.S. states and 19 nations considered in the main paper. Furthermore, it presents additional graphs illustrating the $\mathcal{R}_{0}  $ estimates to facilitate the understanding of their clustering and spatial distribution patterns. It also reports the estimated $\mathcal{R}_{0}$ without any mitigating factors to showcase the biases in the resulting estimates.

First, Table \ref{tab:state_regressors} reports the estimates on the mitigating factors in the panel regressions for the 48 contiguous U.S. states. It complements the $\mathcal{R}_{0}$ (or intercept) estimates shown in Table \ref{tab:state_R0} of the main paper. The values of the multiplication factor (MF) are assumed to decline linearly from 5 to 2 and from 8 to 2.5. The summary statistics of the mitigating factors are given in Tables \ref{tab:state_summary_statistics_subsample} and \ref {tab:state_summary_statistics_fullsample}. The results reveal that the stringency index is negatively associated with the transmission rate, and the association is highly significant for the full sample period. In states with a Democrat governor, economic support is found to contribute to the slowdown of transmission in most cases, although its effect is statistically and practically insignificant. In comparison, having a Republican governor tends to adversely influence the effectiveness of economic support in reducing disease transmission. In addition, a higher proportion of the fully vaccinated population is found to significantly lower the transmission rate, while a larger share of the Delta variant leads to a significant increase in disease spread, as expected. The threshold value reflects when individuals begin to take precautionary measures due to the fear of getting infected with the increasing number of cases. The low threshold values (0.7 for the high-MF pre-vaccination sample and 0.4 for all other cases) suggest that voluntary social distancing took place soon after the outbreak, and it significantly lowers the transmission rate, as evidenced by the large negative estimates on the threshold variable. Changing the MF specifications have minimal impact on our findings.

Turning next to the country-level regressions, we provide the coefficient estimates for the mitigating factors in Table \ref{tab:country_regressors}, which complements the intercept estimates given in Table \ref{tab:country_R0} of the main paper. The summary statistics of the mitigating factors can be found in Tables \ref{tab:country_summary_statistics_subsample} and \ref {tab:country_summary_statistics_fullsample}. Both the stringency index and economic support index are found to significantly reduce Covid-19 transmission, with the stringency policies having a greater effect. The threshold effect is also negative and highly significant, reflecting a considerable impact of precautionary behaviors on disease propagation. In all cases, the threshold parameter is estimated to be 0.01, indicating that the voluntary behavioral changes occurred shortly after the outbreak, which is consistent with our finding for the U.S. In addition, the full-sample results suggest that the share of the population fully vaccinated has a negative and strongly significant effect in decreasing the spread of the disease. The estimates on the share of the Delta variant are positive although insignificant. Their lack of accuracy is likely due to the poor data quality on the Delta variant in less developed countries in our sample. Overall, altering the MF assumptions produces negligible changes in the coefficients on all factors.

\begin{figure}[tbh]
\caption{Estimates of $\mathcal{R}_{0}$ across 48 contiguous U.S. states, pre-vaccination sample (ending January 31, 2021)}
\label{fig:us_r0_subsample}\centering
\begin{subfigure}{0.8\textwidth}
\caption{Multiplication Factor 5 to 2}
\includegraphics[width=\linewidth]{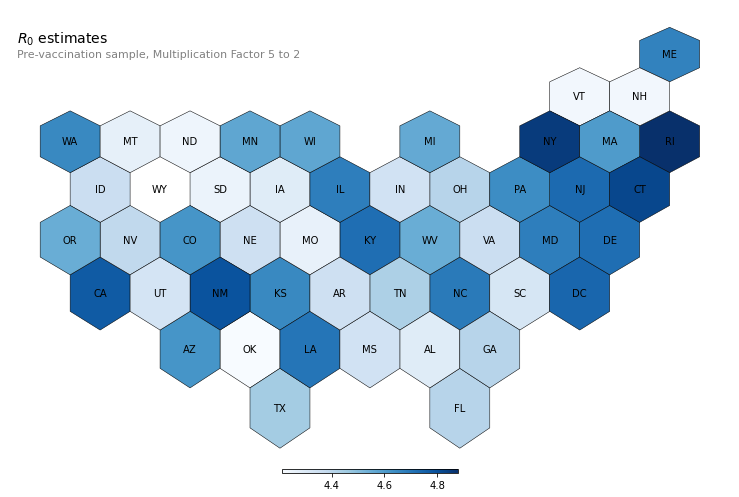}
\label{fig:subsample_states_5_2}
\end{subfigure}
\par
\begin{subfigure}{0.8\textwidth}
\caption{Multiplication Factor 8 to 2.5}
\includegraphics[width=\linewidth]{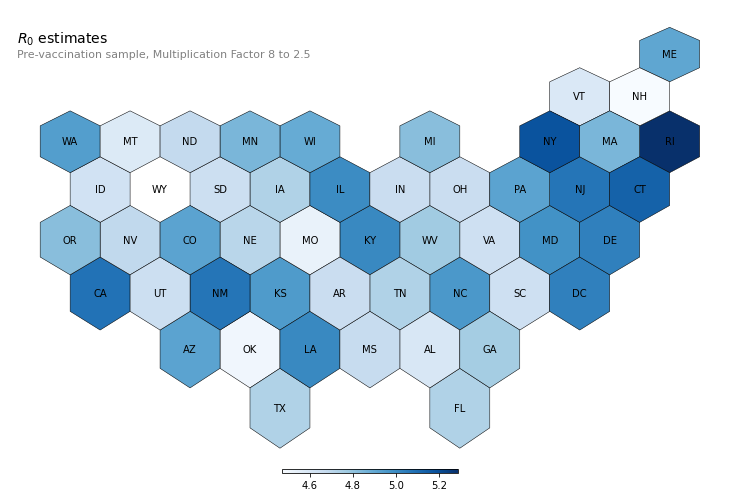}
\label{fig:subsample_states_8_25}
\end{subfigure}
\end{figure}

\begin{figure}[tbh]
    \caption{Estimates of $\mathcal{R}_{0}$ across 48 contiguous U.S. states, full sample (ending November 30, 2021)}
    \label{fig:us_r0_full_sample}\centering
    \begin{subfigure}{0.8\textwidth}
        \caption{Multiplication Factor 5 to 2}
        \includegraphics[width=\linewidth]{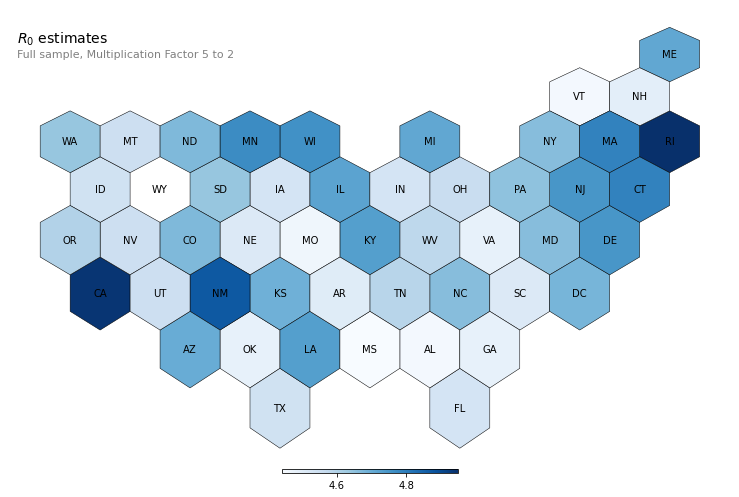}
        \label{fig:full_sample_states_5_2}
    \end{subfigure}
    \par
    \begin{subfigure}{0.8\textwidth}
        \caption{Multiplication Factor 8 to 2.5}
        \includegraphics[width=\linewidth]{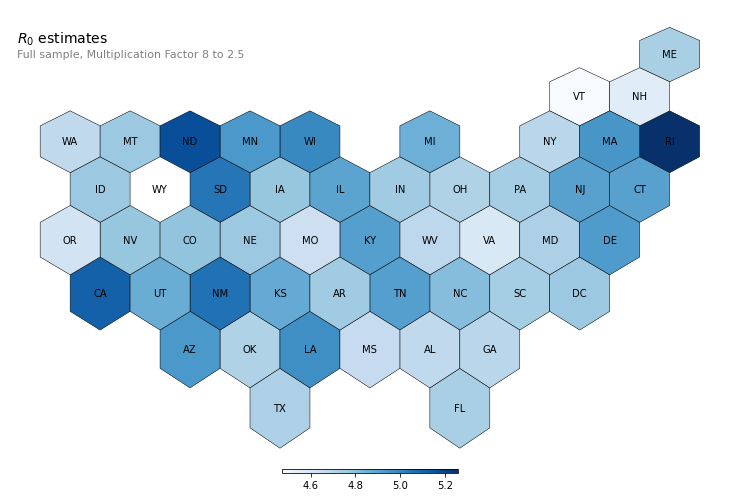}
        \label{fig:full_sample_states_8_25}
    \end{subfigure}
\end{figure}

\begin{figure}[thb!]
    \caption{Estimates of $\mathcal{R}_0$ across 19 countries, pre-vaccination sample (ending January 31, 2021)}
    \label{fig:int_r0_subsample}\centering
    \begin{subfigure}{0.8\textwidth}
        \caption{Multiplication Factor 5 to 2}
        \includegraphics[width=\linewidth]{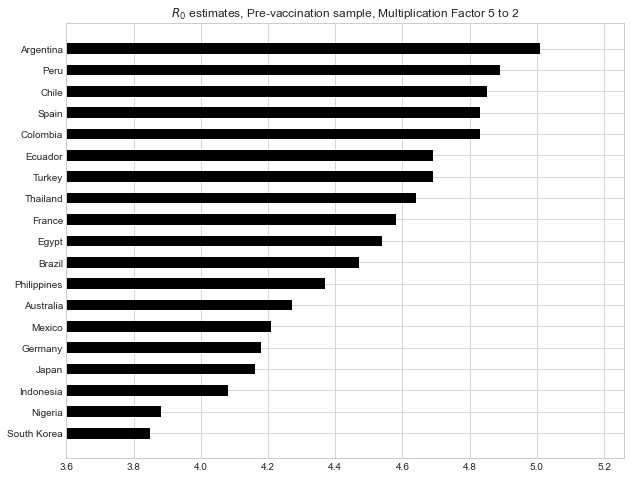}
        \label{fig:subsample_int_5_2}
    \end{subfigure}
    \par
    \begin{subfigure}{0.8\textwidth}
        \caption{Multiplication Factor 8 to 2.5}
        \includegraphics[width=\linewidth]{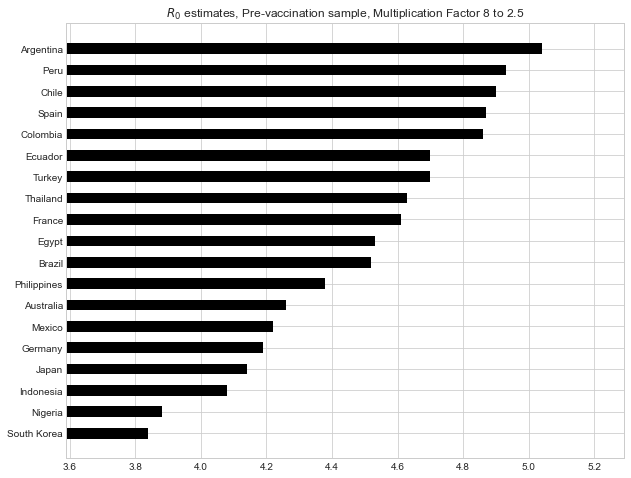}
        \label{fig:subsample_int_8_25}
    \end{subfigure}
\end{figure}

\begin{figure}[thb!]
    \caption{Estimates of $\mathcal{R}_0$ across 19 countries, full sample (ending November 30, 2021)}
    \label{fig:int_r0_full_sample}\centering
    \begin{subfigure}{0.8\textwidth}
        \caption{Multiplication Factor 5 to 2}
        \includegraphics[width=\linewidth]{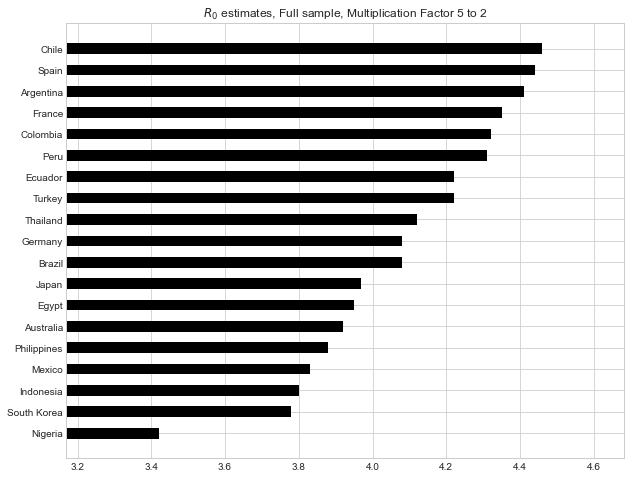}
        \label{fig:full_sample_int_5_2}
    \end{subfigure}
    \par
    \begin{subfigure}{0.8\textwidth}
        \caption{Multiplication Factor 8 to 2.5}
        \includegraphics[width=\linewidth]{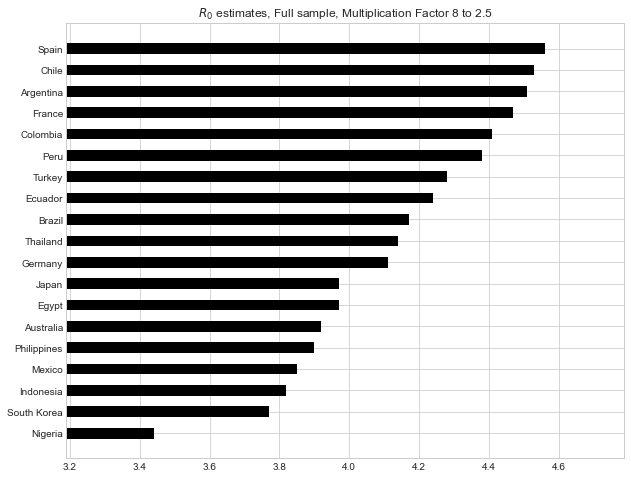}
        \label{fig:rull_sample_int_8_25}
    \end{subfigure}
\end{figure}

In addition, we provide supplementary figures to illustrate the distributions of the $\mathcal{R}_{0}$ estimates presented in Tables \ref {tab:state_R0} and \ref{tab:country_R0} of the main paper. Hexagon maps, displayed in Figures \ref{fig:us_r0_subsample} and \ref {fig:us_r0_full_sample}, have been included to illustrate the spatial distribution of the $\mathcal{R}_{0}$ estimates across U.S. states. To visualize the country-specific $\mathcal{R}_{0}$ estimates, Figures \ref {fig:int_r0_subsample} and \ref{fig:int_r0_full_sample} display the estimates in descending order for different sample periods and MF specifications.

Lastly, as discussed in Section \ref{subsec:est_without_mitigating} of the main paper, we provide the $\mathcal{R}_{0}$ estimates without mitigating factors in Tables \ref{tab:state_R0_no_mit} and \ref{tab:county_R0_no_mit} for the U.S. states and countries, respectively. These tables serve to demonstrate the substantial underestimation of $\mathcal{R}_{0}$ resulting from disregarding mitigating factors.

\begin{table}[!th]\caption{Estimation results for the mitigating factors across 48 contiguous U.S. states}\centering 
\begin{small}
\begin{tabular}{lccccc}\toprule 
&\multicolumn{2}{c}{\bf Pre-vaccination sample}&&\multicolumn{2}{c}{\bf Full sample} \\
&\multicolumn{2}{c}{\bf Ending Jan 31, 2021}&&\multicolumn{2}{c}{\bf Ending Nov 30, 2021} \\
\cline{2-3} \cline{3-3} \cline{5-6} \cline{6-6} 
& MF & MF && MF& MF \\
State& 5 to 2& 8 to 2.5 && 5 to 2 & 8 to 2.5 \\ \midrule
\bf Stringency Index&-1.28&-1.27&&-1.07&-1.51\\ 
standard s.e. (t-ratio)&0.08 (-15.4)&0.09 (-14.5)&&0.11 (-9.6)&0.13 (-11.5)\\ 
robust1 s.e. (t-ratio)&0.15 (-8.7)&0.14 (-9.1)&&0.14 (-7.7)&0.15 (-9.8)\\ 
robust2 s.e. (t-ratio)&1.16 (-1.1)&0.97 (-1.3)&&0.64 (-1.7)&0.60 (-2.5)\\ \midrule 
\bf Economic Support&-0.39&-0.63&&-0.03&0.02\\ 
standard s.e. (t-ratio)&0.06 (-6.3)&0.06 (-9.9)&&0.07 (-0.4)&0.09 (0.3)\\ 
robust1 s.e. (t-ratio)&0.12 (-3.2)&0.12 (-5.2)&&0.09 (-0.3)&0.10 (0.2)\\ 
robust2 s.e. (t-ratio)&0.32 (-1.2)&0.33 (-1.9)&&0.20 (-0.1)&0.23 (0.1)\\ \midrule 
\bf Vaccinated Share&&&&-1.65&-1.51\\ 
standard s.e. (t-ratio)&-&-&&0.10 (-17.2)&0.11 (-13.3)\\ 
robust1 s.e. (t-ratio)&-&-&&0.08 (-19.8)&0.10 (-15.2)\\ 
robust2 s.e. (t-ratio)&-&-&&0.40 (-4.2)&0.44 (-3.4)\\ \midrule 
\bf Delta Variant Share&&&&1.00&1.00\\ 
standard s.e. (t-ratio)&-&-&&0.04 (22.8)&0.05 (19.4)\\ 
robust1 s.e. (t-ratio)&-&-&&0.04 (24.7)&0.05 (18.9)\\ 
robust2 s.e. (t-ratio)&-&-&&0.18 (5.5)&0.24 (4.1)\\ \midrule 
\bf Rep $\times$ Economic Support&0.42&0.39&&0.26&0.18\\ 
standard s.e. (t-ratio)&0.07 (6.1)&0.07 (5.4)&&0.09 (3.0)&0.10 (1.8)\\ 
robust1 s.e. (t-ratio)&0.15 (2.8)&0.14 (2.7)&&0.10 (2.7)&0.11 (1.7)\\ 
robust2 s.e. (t-ratio)&0.14 (3.0)&0.14 (2.7)&&0.09 (2.8)&0.10 (1.8)\\ \midrule 
\bf Threshold Indicator&-2.31&-2.37&&-2.71&-2.49\\ 
standard s.e. (t-ratio)&0.04 (-62.2)&0.04 (-54.6)&&0.07 (-39.1)&0.08 (-30.6)\\ 
robust1 s.e. (t-ratio)&0.11 (-20.3)&0.12 (-19.1)&&0.13 (-20.7)&0.13 (-19.3)\\ 
robust2 s.e. (t-ratio)&0.50 (-4.6)&0.53 (-4.5)&&0.59 (-4.6)&0.55 (-4.5)\\ \midrule 
Threshold Value&0.70&0.40&&0.40&0.40\\ 
R-squared&0.31&0.29&&0.09&0.08\\ 
\bottomrule\end{tabular}
\end{small}
\begin{tablenotes}
\footnotesize
\item
Notes: The pre-vaccination sample includes 14,996 observations for 48 contiguous states, covering a time span of $T_{\min}=163$ to $T_{\max}=331$ days. The full sample comprises 29,531 observations, spanning $T_{\min}=457$  to $T_{\max}=634$ days. Both estimation samples are unbalanced at the beginning.
Rep is a dummy variable equal to one for a Republican governor.
``Robust1" standard errors are robust to serial correlation only (Newey-West type correction), whereas ``robust2" standard errors are robust to serial correlation as well as any cross-sectional correlation. 
Figures in parentheses are t-ratios. Oxford stringency and economic support indices are divided by 100 so that they take values between zero and one.
Lag order is set to $p$ = 10 days in all regressions. The multiplication factor (MF) declines linearly from the high value to the low value.
\end{tablenotes}
\label{tab:state_regressors}
\end{table} 
\begin{table}
\centering
\caption{Summary statistics by state for the pre-vaccination sample (ending January 31, 2021)}
{\footnotesize
\begin{tabular}{l|cc|cc|cc|cc|cc|}
\toprule
{} & \multicolumn{2}{c}{stringency} & \multicolumn{2}{c}{support} & \multicolumn{2}{c}{thr} \\
{} &       mean &    std &    mean &    std & mean &    std \\
\midrule
\textbf{Alabama             } &       0.52 & (0.12) &    0.43 & (0.18) & 0.97 & (0.18) \\
\textbf{Arizona             } &       0.58 & (0.07) &    0.47 & (0.14) & 0.97 & (0.18) \\
\textbf{Arkansas            } &       0.57 & (0.09) &    0.44 & (0.18) & 0.97 & (0.16) \\
\textbf{California          } &       0.65 & (0.12) &    0.70 & (0.16) & 0.93 & (0.25) \\
\textbf{Colorado            } &       0.62 & (0.12) &    0.70 & (0.18) & 0.97 & (0.18) \\
\textbf{Connecticut         } &       0.69 & (0.08) &    0.76 & (0.16) & 0.97 & (0.17) \\
\textbf{Delaware            } &       0.67 & (0.12) &    0.71 & (0.16) & 0.98 & (0.13) \\
\textbf{District of Columbia} &       0.73 & (0.08) &    0.68 & (0.18) & 0.99 & (0.10) \\
\textbf{Florida             } &       0.59 & (0.17) &    0.57 & (0.27) & 0.95 & (0.21) \\
\textbf{Georgia             } &       0.61 & (0.11) &    0.64 & (0.25) & 0.96 & (0.20) \\
\textbf{Idaho               } &       0.61 & (0.12) &    0.40 & (0.21) & 0.97 & (0.17) \\
\textbf{Illinois            } &       0.62 & (0.14) &    0.61 & (0.24) & 0.96 & (0.19) \\
\textbf{Indiana             } &       0.57 & (0.09) &    0.59 & (0.25) & 0.96 & (0.19) \\
\textbf{Iowa                } &       0.53 & (0.10) &    0.81 & (0.15) & 0.97 & (0.17) \\
\textbf{Kansas              } &       0.61 & (0.08) &    0.51 & (0.23) & 0.97 & (0.17) \\
\textbf{Kentucky            } &       0.69 & (0.11) &    0.61 & (0.25) & 0.97 & (0.18) \\
\textbf{Louisiana           } &       0.64 & (0.11) &    0.50 & (0.25) & 0.97 & (0.17) \\
\textbf{Maine               } &       0.76 & (0.12) &    0.60 & (0.24) & 0.99 & (0.10) \\
\textbf{Maryland            } &       0.67 & (0.15) &    0.68 & (0.19) & 0.96 & (0.19) \\
\textbf{Massachusetts       } &       0.64 & (0.11) &    0.42 & (0.19) & 0.96 & (0.20) \\
\textbf{Michigan            } &       0.64 & (0.15) &    0.44 & (0.17) & 0.96 & (0.19) \\
\textbf{Minnesota           } &       0.63 & (0.09) &    0.46 & (0.14) & 0.97 & (0.18) \\
\textbf{Mississippi         } &       0.56 & (0.13) &    0.48 & (0.24) & 0.97 & (0.16) \\
\textbf{Missouri            } &       0.53 & (0.12) &    0.42 & (0.18) & 0.97 & (0.18) \\
\textbf{Montana             } &       0.57 & (0.07) &    0.41 & (0.20) & 0.97 & (0.17) \\
\textbf{Nebraska            } &       0.59 & (0.10) &    0.58 & (0.15) & 0.97 & (0.16) \\
\textbf{Nevada              } &       0.61 & (0.09) &    0.50 & (0.23) & 0.97 & (0.17) \\
\textbf{New Hampshire       } &       0.57 & (0.16) &    0.50 & (0.29) & 0.98 & (0.13) \\
\textbf{New Jersey          } &       0.62 & (0.13) &    0.70 & (0.18) & 0.97 & (0.18) \\
\textbf{New Mexico          } &       0.81 & (0.04) &    0.48 & (0.12) & 0.98 & (0.15) \\
\textbf{New York            } &       0.72 & (0.11) &    0.95 & (0.22) & 0.95 & (0.23) \\
\textbf{North Carolina      } &       0.67 & (0.08) &    0.59 & (0.10) & 0.96 & (0.21) \\
\textbf{North Dakota        } &       0.47 & (0.05) &    0.41 & (0.20) & 1.00 & (0.00) \\
\textbf{Ohio                } &       0.64 & (0.09) &    0.46 & (0.15) & 0.96 & (0.21) \\
\textbf{Oklahoma            } &       0.47 & (0.14) &    0.39 & (0.21) & 0.97 & (0.18) \\
\textbf{Oregon              } &       0.64 & (0.09) &    0.71 & (0.16) & 0.97 & (0.18) \\
\textbf{Pennsylvania        } &       0.63 & (0.11) &    0.63 & (0.19) & 0.96 & (0.20) \\
\textbf{Rhode Island        } &       0.70 & (0.07) &    0.86 & (0.07) & 0.99 & (0.10) \\
\textbf{South Carolina      } &       0.54 & (0.11) &    0.51 & (0.19) & 0.97 & (0.18) \\
\textbf{South Dakota        } &       0.47 & (0.08) &    0.43 & (0.17) & 0.99 & (0.08) \\
\textbf{Tennessee           } &       0.61 & (0.09) &    0.42 & (0.19) & 0.97 & (0.18) \\
\textbf{Texas               } &       0.59 & (0.10) &    0.43 & (0.18) & 0.95 & (0.21) \\
\textbf{Utah                } &       0.54 & (0.07) &    0.39 & (0.21) & 0.97 & (0.17) \\
\textbf{Vermont             } &       0.69 & (0.09) &    0.67 & (0.17) & 0.96 & (0.19) \\
\textbf{Virginia            } &       0.61 & (0.10) &    0.57 & (0.21) & 0.96 & (0.20) \\
\textbf{Washington          } &       0.66 & (0.12) &    0.75 & (0.19) & 0.97 & (0.18) \\
\textbf{West Virginia       } &       0.62 & (0.09) &    0.49 & (0.14) & 0.98 & (0.14) \\
\textbf{Wisconsin           } &       0.58 & (0.11) &    0.52 & (0.21) & 0.97 & (0.18) \\
\textbf{All                 } &       0.62 & (0.13) &    0.56 & (0.24) & 0.97 & (0.18) \\
\bottomrule
\end{tabular}
}
\label{tab:state_summary_statistics_subsample}
\end{table}
 
\begin{table}
\centering
\caption{Summary statistics by state for the full sample (ending November 30, 2021)}
{\footnotesize
\begin{tabular}{l|cc|cc|cc|cc|cc|}
\toprule
{} & \multicolumn{2}{c}{delta} & \multicolumn{2}{c}{stringency} & \multicolumn{2}{c}{support} & \multicolumn{2}{c}{thr} & \multicolumn{2}{c}{vaxx} \\
{} &  mean &    std &       mean &    std &    mean &    std & mean &    std & mean &    std \\
\midrule
\textbf{Alabama             } &  0.26 & (0.41) &       0.43 & (0.13) &    0.39 & (0.14) & 0.98 & (0.13) & 0.13 & (0.17) \\
\textbf{Arizona             } &  0.25 & (0.41) &       0.46 & (0.16) &    0.41 & (0.12) & 0.98 & (0.13) & 0.17 & (0.21) \\
\textbf{Arkansas            } &  0.28 & (0.43) &       0.48 & (0.13) &    0.41 & (0.15) & 0.99 & (0.11) & 0.14 & (0.18) \\
\textbf{California          } &  0.26 & (0.41) &       0.55 & (0.17) &    0.67 & (0.12) & 0.96 & (0.19) & 0.19 & (0.24) \\
\textbf{Colorado            } &  0.27 & (0.42) &       0.52 & (0.16) &    0.55 & (0.24) & 0.98 & (0.13) & 0.20 & (0.25) \\
\textbf{Connecticut         } &  0.25 & (0.41) &       0.55 & (0.19) &    0.67 & (0.21) & 0.98 & (0.13) & 0.23 & (0.29) \\
\textbf{Delaware            } &  0.25 & (0.41) &       0.55 & (0.16) &    0.62 & (0.17) & 0.99 & (0.09) & 0.19 & (0.24) \\
\textbf{District of Columbia} &  0.24 & (0.41) &       0.59 & (0.17) &    0.65 & (0.15) & 1.00 & (0.07) & 0.20 & (0.25) \\
\textbf{Florida             } &  0.24 & (0.41) &       0.46 & (0.20) &    0.44 & (0.27) & 0.98 & (0.15) & 0.18 & (0.23) \\
\textbf{Georgia             } &  0.25 & (0.41) &       0.50 & (0.16) &    0.49 & (0.29) & 0.98 & (0.15) & 0.14 & (0.18) \\
\textbf{Idaho               } &  0.24 & (0.41) &       0.49 & (0.16) &    0.39 & (0.16) & 0.99 & (0.12) & 0.14 & (0.18) \\
\textbf{Illinois            } &  0.25 & (0.42) &       0.51 & (0.18) &    0.59 & (0.18) & 0.98 & (0.14) & 0.18 & (0.23) \\
\textbf{Indiana             } &  0.26 & (0.42) &       0.48 & (0.14) &    0.47 & (0.22) & 0.98 & (0.14) & 0.16 & (0.20) \\
\textbf{Iowa                } &  0.27 & (0.42) &       0.43 & (0.14) &    0.58 & (0.28) & 0.99 & (0.12) & 0.19 & (0.23) \\
\textbf{Kansas              } &  0.28 & (0.43) &       0.49 & (0.15) &    0.53 & (0.18) & 0.99 & (0.12) & 0.17 & (0.21) \\
\textbf{Kentucky            } &  0.24 & (0.41) &       0.56 & (0.18) &    0.48 & (0.22) & 0.98 & (0.13) & 0.17 & (0.21) \\
\textbf{Louisiana           } &  0.24 & (0.41) &       0.58 & (0.10) &    0.42 & (0.20) & 0.98 & (0.13) & 0.15 & (0.18) \\
\textbf{Maine               } &  0.23 & (0.41) &       0.58 & (0.21) &    0.49 & (0.21) & 1.00 & (0.07) & 0.23 & (0.29) \\
\textbf{Maryland            } &  0.24 & (0.41) &       0.54 & (0.19) &    0.64 & (0.21) & 0.98 & (0.14) & 0.21 & (0.27) \\
\textbf{Massachusetts       } &  0.25 & (0.41) &       0.56 & (0.16) &    0.37 & (0.18) & 0.98 & (0.15) & 0.22 & (0.29) \\
\textbf{Michigan            } &  0.24 & (0.41) &       0.53 & (0.19) &    0.40 & (0.14) & 0.98 & (0.14) & 0.17 & (0.22) \\
\textbf{Minnesota           } &  0.25 & (0.42) &       0.51 & (0.17) &    0.37 & (0.18) & 0.98 & (0.13) & 0.20 & (0.25) \\
\textbf{Mississippi         } &  0.27 & (0.42) &       0.46 & (0.16) &    0.43 & (0.19) & 0.99 & (0.12) & 0.13 & (0.17) \\
\textbf{Missouri            } &  0.29 & (0.44) &       0.45 & (0.14) &    0.40 & (0.15) & 0.98 & (0.14) & 0.16 & (0.20) \\
\textbf{Montana             } &  0.27 & (0.42) &       0.45 & (0.14) &    0.39 & (0.15) & 0.99 & (0.12) & 0.19 & (0.21) \\
\textbf{Nebraska            } &  0.27 & (0.43) &       0.46 & (0.15) &    0.64 & (0.22) & 0.98 & (0.14) & 0.19 & (0.23) \\
\textbf{Nevada              } &  0.27 & (0.42) &       0.51 & (0.16) &    0.39 & (0.23) & 0.99 & (0.12) & 0.17 & (0.21) \\
\textbf{New Hampshire       } &  0.25 & (0.41) &       0.48 & (0.15) &    0.44 & (0.22) & 0.99 & (0.09) & 0.21 & (0.26) \\
\textbf{New Jersey          } &  0.26 & (0.42) &       0.51 & (0.17) &    0.66 & (0.15) & 0.98 & (0.13) & 0.21 & (0.27) \\
\textbf{New Mexico          } &  0.25 & (0.41) &       0.63 & (0.21) &    0.43 & (0.12) & 0.99 & (0.11) & 0.21 & (0.26) \\
\textbf{New York            } &  0.25 & (0.41) &       0.58 & (0.20) &    0.84 & (0.29) & 0.97 & (0.17) & 0.20 & (0.26) \\
\textbf{North Carolina      } &  0.25 & (0.41) &       0.56 & (0.16) &    0.50 & (0.22) & 0.98 & (0.15) & 0.16 & (0.20) \\
\textbf{North Dakota        } &  0.25 & (0.42) &       0.40 & (0.09) &    0.46 & (0.18) & 1.00 & (0.00) & 0.16 & (0.19) \\
\textbf{Ohio                } &  0.25 & (0.41) &       0.52 & (0.16) &    0.40 & (0.13) & 0.98 & (0.15) & 0.17 & (0.21) \\
\textbf{Oklahoma            } &  0.27 & (0.42) &       0.44 & (0.11) &    0.38 & (0.17) & 0.98 & (0.13) & 0.16 & (0.19) \\
\textbf{Oregon              } &  0.23 & (0.40) &       0.55 & (0.16) &    0.65 & (0.13) & 0.98 & (0.13) & 0.20 & (0.25) \\
\textbf{Pennsylvania        } &  0.25 & (0.42) &       0.49 & (0.18) &    0.56 & (0.16) & 0.98 & (0.14) & 0.19 & (0.24) \\
\textbf{Rhode Island        } &  0.25 & (0.41) &       0.57 & (0.17) &    0.70 & (0.19) & 1.00 & (0.07) & 0.23 & (0.28) \\
\textbf{South Carolina      } &  0.25 & (0.41) &       0.46 & (0.13) &    0.40 & (0.22) & 0.98 & (0.13) & 0.15 & (0.19) \\
\textbf{South Dakota        } &  0.27 & (0.43) &       0.40 & (0.11) &    0.39 & (0.13) & 1.00 & (0.06) & 0.19 & (0.22) \\
\textbf{Tennessee           } &  0.25 & (0.41) &       0.50 & (0.16) &    0.43 & (0.14) & 0.98 & (0.13) & 0.15 & (0.18) \\
\textbf{Texas               } &  0.25 & (0.41) &       0.49 & (0.13) &    0.41 & (0.13) & 0.98 & (0.15) & 0.16 & (0.21) \\
\textbf{Utah                } &  0.28 & (0.43) &       0.44 & (0.12) &    0.38 & (0.16) & 0.99 & (0.12) & 0.16 & (0.20) \\
\textbf{Vermont             } &  0.33 & (0.45) &       0.51 & (0.21) &    0.69 & (0.21) & 0.99 & (0.11) & 0.30 & (0.30) \\
\textbf{Virginia            } &  0.25 & (0.41) &       0.50 & (0.16) &    0.55 & (0.20) & 0.98 & (0.14) & 0.20 & (0.25) \\
\textbf{Washington          } &  0.25 & (0.41) &       0.57 & (0.17) &    0.69 & (0.20) & 0.98 & (0.13) & 0.20 & (0.26) \\
\textbf{West Virginia       } &  0.24 & (0.41) &       0.49 & (0.17) &    0.44 & (0.13) & 0.99 & (0.10) & 0.15 & (0.17) \\
\textbf{Wisconsin           } &  0.25 & (0.41) &       0.49 & (0.14) &    0.41 & (0.21) & 0.98 & (0.13) & 0.19 & (0.24) \\
\textbf{All                 } &  0.26 & (0.42) &       0.51 & (0.17) &    0.50 & (0.22) & 0.98 & (0.13) & 0.18 & (0.23) \\
\bottomrule
\end{tabular}
}
\label{tab:state_summary_statistics_fullsample}
\end{table}

\begin{table}[!t]\caption{Estimation results for the mitigating factors across 19 countries}\centering 
\begin{small}
\begin{tabular}{lccccc}\toprule 
&\multicolumn{2}{c}{\bf Pre-vaccination sample}&&\multicolumn{2}{c}{\bf Full sample} \\
&\multicolumn{2}{c}{\bf Ending Jan 31, 2021}&&\multicolumn{2}{c}{\bf Ending Nov 30, 2021} \\
\cline{2-3} \cline{3-3} \cline{5-6} \cline{6-6} 
& MF & MF && MF& MF \\
Country& 5 to 2& 8 to 2.5 && 5 to 2 & 8 to 2.5 \\ \midrule
\bf Stringency Index&-1.68&-1.71&&-1.04&-1.07\\ 
standard s.e. (t-ratio)&0.14 (-11.8)&0.14 (-11.9)&&0.09 (-11.1)&0.10 (-10.8)\\ 
robust1 s.e. (t-ratio)&0.26 (-6.5)&0.26 (-6.7)&&0.19 (-5.6)&0.19 (-5.7)\\ 
robust2 s.e. (t-ratio)&0.35 (-4.7)&0.35 (-4.9)&&0.32 (-3.2)&0.31 (-3.4)\\ \midrule 
\bf Economic Support&-1.03&-1.01&&-0.61&-0.63\\ 
standard s.e. (t-ratio)&0.09 (-11.9)&0.09 (-11.6)&&0.05 (-11.1)&0.06 (-10.8)\\ 
robust1 s.e. (t-ratio)&0.13 (-7.8)&0.13 (-7.7)&&0.09 (-6.4)&0.09 (-6.6)\\ 
robust2 s.e. (t-ratio)&0.15 (-6.8)&0.15 (-6.8)&&0.14 (-4.4)&0.13 (-4.7)\\ \midrule 
\bf Vaccinated Share&&&&-0.78&-0.62\\ 
standard s.e. (t-ratio)&-&-&&0.09 (-8.6)&0.09 (-6.6)\\ 
robust1 s.e. (t-ratio)&-&-&&0.12 (-6.6)&0.12 (-5.0)\\ 
robust2 s.e. (t-ratio)&-&-&&0.18 (-4.3)&0.19 (-3.4)\\ \midrule 
\bf Delta Variant Share&&&&0.12&0.07\\ 
standard s.e. (t-ratio)&-&-&&0.06 (2.2)&0.06 (1.2)\\ 
robust1 s.e. (t-ratio)&-&-&&0.07 (1.7)&0.08 (1.0)\\ 
robust2 s.e. (t-ratio)&-&-&&0.12 (1.0)&0.12 (0.6)\\ \midrule 
\bf Threshold Indicator&-1.44&-1.42&&-1.81&-1.79\\ 
standard s.e. (t-ratio)&0.12 (-12.2)&0.12 (-11.9)&&0.09 (-19.1)&0.10 (-18.0)\\ 
robust1 s.e. (t-ratio)&0.31 (-4.7)&0.30 (-4.7)&&0.32 (-5.6)&0.32 (-5.6)\\ 
robust2 s.e. (t-ratio)&0.24 (-6.0)&0.24 (-5.9)&&0.35 (-5.2)&0.34 (-5.2)\\ \midrule 
Threshold Value&0.01&0.01&&0.01&0.01\\ 
R-squared&0.13&0.13&&0.09&0.08\\ 
\bottomrule\end{tabular}
\end{small}
\begin{tablenotes}
    \footnotesize
    \item
    Notes: The pre-vaccination sample includes 5,928 observations for 19 countries, covering a time span of $T_{\min}=140$  to $T_{\max}=329$  days. The full sample comprises 11,661 observations for the same 19 countries, spanning $T_{\min}=443$  to $T_{\max}=632$  days. Both estimation samples are unbalanced at the beginning. 
    ``Robust1" standard errors are robust to serial correlation only (Newey-West type correction), whereas ``robust2" standard errors are robust to serial correlation as well as any cross-sectional correlation. 
    Figures in parentheses are t-ratios. Oxford stringency and economic support indices are divided by 100 so that they take values between zero and one.
    Lag order is set to $p$ = 10 days in all regressions. The multiplication factor (MF) declines linearly from the high value to the low value.
\end{tablenotes}
\label{tab:country_regressors}
\end{table} 
\begin{table}
\centering
\caption{Summary statistics by country for the pre-vaccination sample (ending January 31, 2021)}
{\footnotesize
\begin{tabular}{l|cc|cc|cc}
\toprule
{} & \multicolumn{2}{c}{Stringency} & \multicolumn{2}{c}{Support} & \multicolumn{2}{c}{Threshold Ind.} \\
{} &       mean &    std &    mean &    std & mean &    std \\
\midrule
\textbf{Argentina  } &       0.86 & (0.11) &    0.72 & (0.15) & 0.99 & (0.08) \\
\textbf{Australia  } &       0.65 & (0.12) &    0.69 & (0.17) & 1.00 & (0.00) \\
\textbf{Brazil     } &       0.68 & (0.13) &    0.42 & (0.17) & 0.97 & (0.16) \\
\textbf{Chile      } &       0.79 & (0.15) &    0.71 & (0.31) & 0.99 & (0.11) \\
\textbf{Colombia   } &       0.76 & (0.13) &    0.72 & (0.12) & 0.98 & (0.12) \\
\textbf{Ecuador    } &       0.71 & (0.17) &    0.72 & (0.12) & 0.96 & (0.21) \\
\textbf{Egypt      } &       0.72 & (0.15) &    0.68 & (0.17) & 0.98 & (0.14) \\
\textbf{France     } &       0.64 & (0.17) &    0.68 & (0.24) & 0.97 & (0.17) \\
\textbf{Germany    } &       0.62 & (0.13) &    0.42 & (0.16) & 0.98 & (0.13) \\
\textbf{Indonesia  } &       0.63 & (0.09) &    0.28 & (0.10) & 0.96 & (0.19) \\
\textbf{Japan      } &       0.37 & (0.07) &    0.85 & (0.26) & 0.98 & (0.13) \\
\textbf{Mexico     } &       0.70 & (0.15) &    0.25 & (0.35) & 0.98 & (0.15) \\
\textbf{Nigeria    } &       0.68 & (0.14) &    0.15 & (0.19) & 0.93 & (0.26) \\
\textbf{Peru       } &       0.81 & (0.13) &    0.68 & (0.16) & 0.98 & (0.14) \\
\textbf{Philippines} &       0.76 & (0.15) &    0.42 & (0.28) & 0.97 & (0.17) \\
\textbf{South Korea} &       0.56 & (0.09) &    0.45 & (0.13) & 0.98 & (0.13) \\
\textbf{Spain      } &       0.66 & (0.15) &    0.81 & (0.21) & 0.98 & (0.13) \\
\textbf{Thailand   } &       0.60 & (0.16) &    0.81 & (0.39) & 0.84 & (0.37) \\
\textbf{Turkey     } &       0.64 & (0.12) &    0.75 & (0.26) & 0.97 & (0.16) \\
\textbf{All        } &       0.68 & (0.17) &    0.58 & (0.30) & 0.97 & (0.16) \\
\bottomrule
\end{tabular}
}
\label{tab:country_summary_statistics_subsample}
\end{table}
 
\begin{table}
\centering
\caption{Summary statistics by country for the full sample (ending November 30, 2021)}
{\footnotesize
\begin{tabular}{l|cc|cc|cc|cc|cc}
\toprule
{} & \multicolumn{2}{c}{Delta} & \multicolumn{2}{c}{Stringency} & \multicolumn{2}{c}{Support} & \multicolumn{2}{c}{Threshold Ind.} & \multicolumn{2}{c}{Vaccinated} \\
{} &  mean &    std &       mean &    std &    mean &    std & mean &    std & mean &    std \\
\midrule
\textbf{Argentina  } &  0.10 & (0.23) &       0.78 & (0.15) &    0.60 & (0.18) & 1.00 & (0.06) & 0.18 & (0.26) \\
\textbf{Australia  } &  0.31 & (0.43) &       0.63 & (0.12) &    0.41 & (0.35) & 1.00 & (0.00) & 0.13 & (0.22) \\
\textbf{Brazil     } &  0.17 & (0.34) &       0.64 & (0.12) &    0.27 & (0.20) & 0.99 & (0.12) & 0.17 & (0.26) \\
\textbf{Chile      } &  0.14 & (0.31) &       0.74 & (0.16) &    0.67 & (0.39) & 0.99 & (0.08) & 0.26 & (0.32) \\
\textbf{Colombia   } &  0.14 & (0.29) &       0.69 & (0.16) &    0.73 & (0.09) & 0.99 & (0.09) & 0.12 & (0.20) \\
\textbf{Ecuador    } &  0.15 & (0.31) &       0.67 & (0.13) &    0.74 & (0.08) & 0.98 & (0.15) & 0.14 & (0.24) \\
\textbf{Egypt      } &  0.21 & (0.39) &       0.58 & (0.18) &    0.72 & (0.13) & 0.99 & (0.10) & 0.02 & (0.05) \\
\textbf{France     } &  0.24 & (0.41) &       0.59 & (0.16) &    0.57 & (0.23) & 0.97 & (0.16) & 0.21 & (0.29) \\
\textbf{Germany    } &  0.25 & (0.41) &       0.62 & (0.16) &    0.40 & (0.12) & 0.99 & (0.10) & 0.21 & (0.28) \\
\textbf{Indonesia  } &  0.29 & (0.43) &       0.66 & (0.08) &    0.33 & (0.09) & 0.98 & (0.14) & 0.07 & (0.13) \\
\textbf{Japan      } &  0.21 & (0.39) &       0.43 & (0.08) &    0.92 & (0.20) & 0.99 & (0.10) & 0.15 & (0.27) \\
\textbf{Mexico     } &  0.24 & (0.40) &       0.60 & (0.17) &    0.26 & (0.27) & 0.99 & (0.11) & 0.13 & (0.19) \\
\textbf{Nigeria    } &  0.26 & (0.41) &       0.57 & (0.15) &    0.08 & (0.15) & 0.96 & (0.19) & 0.01 & (0.01) \\
\textbf{Peru       } &  0.13 & (0.29) &       0.77 & (0.12) &    0.60 & (0.14) & 0.99 & (0.10) & 0.10 & (0.18) \\
\textbf{Philippines} &  0.20 & (0.37) &       0.75 & (0.11) &    0.24 & (0.29) & 0.99 & (0.12) & 0.05 & (0.10) \\
\textbf{South Korea} &  0.23 & (0.39) &       0.54 & (0.09) &    0.44 & (0.12) & 0.99 & (0.10) & 0.15 & (0.26) \\
\textbf{Spain      } &  0.24 & (0.40) &       0.61 & (0.14) &    0.84 & (0.15) & 0.98 & (0.14) & 0.22 & (0.31) \\
\textbf{Thailand   } &  0.34 & (0.42) &       0.58 & (0.11) &    0.72 & (0.25) & 0.95 & (0.22) & 0.13 & (0.21) \\
\textbf{Turkey     } &  0.25 & (0.39) &       0.62 & (0.15) &    0.54 & (0.33) & 0.99 & (0.11) & 0.17 & (0.24) \\
\textbf{All        } &  0.21 & (0.38) &       0.64 & (0.16) &    0.53 & (0.31) & 0.98 & (0.12) & 0.14 & (0.24) \\
\bottomrule
\end{tabular}
}
\label{tab:country_summary_statistics_fullsample}
\end{table}

\begin{table}[!hp]
\caption{Estimates of $\mathcal{R}_0$ across 48 contiguous U.S. states without mitigating factors}
\centering 
\begin{footnotesize}
\begin{tabular}{lccccc}\toprule 
&\multicolumn{2}{c}{\bf Pre-vaccination sample}&&\multicolumn{2}{c}{\bf Full sample} \\
&\multicolumn{2}{c}{\bf Ending Jan 31, 2021}&&\multicolumn{2}{c}{\bf Ending Nov 30, 2021} \\
\cline{2-3} \cline{3-3} \cline{5-6} \cline{6-6} 
& MF & MF && MF& MF \\
State& 5 to 2& 8 to 2.5 && 5 to 2 & 8 to 2.5 \\ \midrule
    Arizona&1.51 [0.07]&1.81 [0.08]&&1.48 [0.11]&1.61 [0.11]\\
Arkansas&1.45 [0.06]&1.73 [0.07]&&1.40 [0.07]&1.52 [0.07]\\
California&1.66 [0.32]&1.86 [0.37]&&1.53 [0.14]&1.60 [0.14]\\
Colorado&1.38 [0.07]&1.54 [0.07]&&1.34 [0.11]&1.41 [0.11]\\
Connecticut&1.38 [0.09]&1.54 [0.09]&&1.41 [0.13]&1.49 [0.13]\\
Delaware&1.39 [0.07]&1.60 [0.08]&&1.34 [0.09]&1.42 [0.09]\\
District of Columbia&1.25 [0.06]&1.35 [0.07]&&1.27 [0.09]&1.34 [0.09]\\
Florida&1.44 [0.09]&1.67 [0.09]&&1.48 [0.13]&1.58 [0.13]\\
Georgia&1.41 [0.08]&1.63 [0.09]&&1.46 [0.11]&1.56 [0.11]\\
Idaho&1.43 [0.07]&1.67 [0.08]&&1.36 [0.10]&1.47 [0.10]\\
Illinois&1.46 [0.09]&1.68 [0.09]&&1.45 [0.13]&1.56 [0.13]\\
Indiana&1.46 [0.08]&1.70 [0.09]&&1.42 [0.12]&1.52 [0.12]\\
Iowa&1.49 [0.07]&1.79 [0.08]&&1.42 [0.09]&1.56 [0.09]\\
Kansas&1.47 [0.06]&1.71 [0.07]&&1.44 [0.08]&1.55 [0.09]\\
Kentucky&1.42 [0.07]&1.63 [0.08]&&1.39 [0.08]&1.47 [0.08]\\
Louisiana&1.43 [0.09]&1.67 [0.10]&&1.47 [0.14]&1.59 [0.14]\\
Maine&1.22 [0.06]&1.27 [0.06]&&1.22 [0.07]&1.24 [0.07]\\
Maryland&1.30 [0.08]&1.41 [0.08]&&1.37 [0.11]&1.43 [0.11]\\
Massachusetts&1.41 [0.10]&1.59 [0.10]&&1.42 [0.14]&1.51 [0.14]\\
Michigan&1.43 [0.09]&1.60 [0.10]&&1.37 [0.13]&1.45 [0.13]\\
Minnesota&1.47 [0.08]&1.68 [0.09]&&1.39 [0.09]&1.48 [0.09]\\
Mississippi&1.42 [0.07]&1.67 [0.08]&&1.41 [0.09]&1.54 [0.09]\\
Missouri&1.45 [0.07]&1.66 [0.08]&&1.38 [0.10]&1.48 [0.10]\\
Montana&1.44 [0.06]&1.69 [0.07]&&1.36 [0.09]&1.47 [0.09]\\
Nebraska&1.47 [0.07]&1.75 [0.08]&&1.38 [0.09]&1.52 [0.09]\\
Nevada&1.42 [0.07]&1.65 [0.07]&&1.39 [0.10]&1.49 [0.10]\\
New Hampshire&1.29 [0.07]&1.38 [0.07]&&1.27 [0.08]&1.30 [0.08]\\
New Jersey&1.44 [0.10]&1.66 [0.11]&&1.46 [0.17]&1.56 [0.17]\\
New Mexico&1.40 [0.06]&1.58 [0.07]&&1.34 [0.08]&1.42 [0.09]\\
New York&1.30 [0.11]&1.37 [0.11]&&1.40 [0.18]&1.44 [0.17]\\
North Carolina&1.37 [0.07]&1.54 [0.07]&&1.41 [0.09]&1.48 [0.09]\\
North Dakota&1.62 [0.08]&2.17 [0.11]&&1.36 [0.07]&1.62 [0.08]\\
Ohio&1.40 [0.08]&1.57 [0.08]&&1.41 [0.11]&1.48 [0.11]\\
Oklahoma&1.44 [0.07]&1.72 [0.08]&&1.40 [0.08]&1.52 [0.08]\\
Oregon&1.23 [0.06]&1.29 [0.06]&&1.26 [0.08]&1.29 [0.08]\\
Pennsylvania&1.39 [0.09]&1.54 [0.09]&&1.40 [0.13]&1.46 [0.13]\\
Rhode Island&1.49 [0.08]&1.84 [0.10]&&1.38 [0.10]&1.52 [0.11]\\
South Carolina&1.44 [0.08]&1.70 [0.09]&&1.44 [0.09]&1.55 [0.09]\\
South Dakota&1.57 [0.08]&2.04 [0.11]&&1.38 [0.09]&1.59 [0.10]\\
Tennessee&1.49 [0.08]&1.83 [0.10]&&1.43 [0.09]&1.56 [0.09]\\
Texas&1.44 [0.10]&1.65 [0.10]&&1.53 [0.15]&1.62 [0.15]\\
Utah&1.52 [0.06]&1.86 [0.07]&&1.43 [0.09]&1.57 [0.09]\\
Vermont&1.21 [0.08]&1.25 [0.08]&&1.26 [0.15]&1.28 [0.15]\\
Virginia&1.33 [0.07]&1.45 [0.08]&&1.40 [0.11]&1.46 [0.11]\\
Washington&1.26 [0.07]&1.33 [0.07]&&1.30 [0.11]&1.34 [0.11]\\
West Virginia&1.35 [0.06]&1.50 [0.07]&&1.31 [0.07]&1.35 [0.07]\\
Wisconsin&1.49 [0.07]&1.77 [0.08]&&1.41 [0.08]&1.54 [0.09]\\
\bottomrule\end{tabular}
\end{footnotesize}
\begin{tablenotes}
\scriptsize
\item
Notes: Numbers in brackets are standard errors robust to serial correlation and cross-sectional correlation (robust2). The multiplication factor (MF) declines linearly from the high value to the low value.
\end{tablenotes}
\label{tab:state_R0_no_mit}
\end{table}

    \begin{table}[!hp]
    \caption{Estimates of $\mathcal{R}_0$ across 19 countries without mitigating factors}
    \centering 
    \begin{footnotesize}
    \begin{tabular}{lccccc}\toprule 
    &\multicolumn{2}{c}{\bf Pre-vaccination sample}&&\multicolumn{2}{c}{\bf Full sample} \\
    &\multicolumn{2}{c}{\bf Ending Jan 31, 2021}&&\multicolumn{2}{c}{\bf Ending Nov 30, 2021} \\
    \cline{2-3} \cline{3-3} \cline{5-6} \cline{6-6} 
    & MF & MF && MF& MF \\
    Country& 5 to 2& 8 to 2.5 && 5 to 2 & 8 to 2.5 \\ \midrule
    Australia&1.14 [0.09]&1.14 [0.09]&&1.04 [0.13]&1.04 [0.13]\\
Brazil&1.45 [0.07]&1.45 [0.07]&&1.54 [0.12]&1.54 [0.12]\\
Chile&1.38 [0.06]&1.38 [0.06]&&1.42 [0.09]&1.42 [0.09]\\
Colombia&1.37 [0.05]&1.37 [0.05]&&1.44 [0.06]&1.44 [0.06]\\
Ecuador&1.23 [0.08]&1.23 [0.08]&&1.39 [0.15]&1.39 [0.15]\\
Egypt&1.12 [0.05]&1.12 [0.05]&&1.22 [0.08]&1.22 [0.08]\\
France&1.69 [0.17]&1.69 [0.17]&&1.55 [0.18]&1.55 [0.18]\\
Germany&1.36 [0.10]&1.36 [0.10]&&1.41 [0.14]&1.41 [0.14]\\
Indonesia&1.12 [0.06]&1.12 [0.06]&&1.34 [0.07]&1.34 [0.07]\\
Japan&1.09 [0.07]&1.09 [0.07]&&1.26 [0.08]&1.26 [0.08]\\
Mexico&1.20 [0.06]&1.20 [0.06]&&1.38 [0.08]&1.38 [0.08]\\
Nigeria&1.07 [0.06]&1.07 [0.06]&&1.25 [0.08]&1.25 [0.08]\\
Peru&1.35 [0.07]&1.35 [0.07]&&1.47 [0.11]&1.47 [0.11]\\
Philippines&1.17 [0.06]&1.17 [0.06]&&1.27 [0.09]&1.27 [0.09]\\
South Korea&1.10 [0.05]&1.10 [0.05]&&1.09 [0.09]&1.09 [0.09]\\
Spain&1.53 [0.12]&1.53 [0.12]&&1.59 [0.18]&1.59 [0.18]\\
Thailand&1.31 [0.14]&1.31 [0.14]&&1.59 [0.37]&1.59 [0.37]\\
Turkey&1.43 [0.12]&1.43 [0.12]&&1.46 [0.22]&1.46 [0.22]\\
    \bottomrule\end{tabular}
    \end{footnotesize}
    \begin{tablenotes}
    \scriptsize
    \item
    Notes: Numbers in brackets are standard errors robust to serial correlation and cross-sectional correlation (robust2). The multiplication factor (MF) declines linearly from the high value to the low value.
    \end{tablenotes}
    \label{tab:county_R0_no_mit}
    \end{table}

\end{document}